\documentclass[preprint,pra,aps,floatfix,showpacs]{revtex4-1}
\usepackage{epsfig,graphicx,tabularx,epstopdf}

\begin{document}
\title{ dc to ac Josephson transition in a dc atom superconducting quantum interference device}
\author{H. M. Cataldo}
\affiliation{IFIBA-CONICET
\\ and \\
Departamento de F\'{\i}sica, FCEN-UBA
Pabell\'on 1, Ciudad Universitaria, 1428 Buenos Aires, Argentina}
\begin{abstract}
We analyze
the effect of the barrier 
motion on the Bose-Hubbard Hamiltonian of a ring-shaped Bose-Einstein condensate
interrupted by a pair of Josephson junctions, a configuration which is the cold atom analog of the well-known dc
superconducting quantum interference device (SQUID). Such an effect is also shown to modify the Heisenberg
equation of motion of the boson field operator in the two-mode approximation, where a hysteretic
contribution that could affect the dynamics for accelerated or overlapping barriers is identified.
By studying the energy landscape as a function of order and control parameters, we
determine the diagram with the location of the dc and ac Josephson regimes, along with the critical points that
are shown to depend on the junctions position. We analyze the dc to ac Josephson transition for adiabatic
barrier trajectories that lead to a final uniform velocity, or which perform symmetric velocity paths.
We show that such symmetric trajectories may induce, when reaching the critical point, highly hysteretic
oscillating return paths within the dc regime, similar to the underdamped hysteresis loops arising from
the action of a resistive flow in the ac regime. We also consider nonequilibrium initial conditions resulting from
a finite phase difference on either side of the junctions, along with the critical features of such a parameter. An excellent agreement between the Gross-Pitaevskii simulations
and the two-mode results is found in all cases.

\end{abstract}
%
%\pacs{03.75.Lm, 03.75.Hh, 03.75.Kk}
%

\maketitle

\section{Introduction}
A direct current 
superconducting quantum interference device (dc-SQUID) basically consists of a ring of superconducting
wire interrupted by two non-superconducting barriers (Josephson junctions). 
Wire leads connected to each side of the device act as a splitter and a recombiner, as
a steady bias current flowing from the splitter  enters the ring and gives rise to the quantum interference of currents emerging from each Josephson junction (JJ) at reaching the recombiner \cite{braginski}. 
A magnetic field threading the ring causes a phase shift between
both currents, an effect which may be utilized to implement a magnetic flux detector. As a result,
dc-SQUIDs constitute today the most sensitive detectors for magnetic flux available \cite{braginski}.
On the other hand,
in a superfluid, the role of the magnetic field is played by rotation, and superfluid helium dc-SQUIDs acting
as rotation sensors have been experimentally implemented \cite{satop}. More recently, 
a cold atom analog of the dc-SQUID was created on a toroidal Bose-Einstein condensate (BEC),
which works as follows \cite{boshier13,sato,sackett}. By slowly moving a pair of JJs  
circumferentially toward each other, there is an induced atom flow through the junctions that keeps the density
and the chemical potential unchanged at both sides of the barriers. 
This is in close analogy to the 
superconducting dc Josephson effect, where a direct current may flow across a JJ without a
driving potential difference \cite{barone,tink,gross2016}. 
However, if the speed of the junctions exceeds certain value such that the induced
atom flow through them reaches the Josephson critical current, the condensate dynamics makes a transition
to the ac Josephson regime, where there is an oscillating current through the JJs but no net current
across them. Therefore, under these conditions the moving barriers simply push the atoms, resulting in
compression of atoms in one sector of the condensate and expansion in the other. Such a difference on
densities yields different values of the chemical potential at both sectors.
Again, this dynamics is analog to the ac Josephson effect in superconductors, where a constant voltage
across a JJ produces an alternating current \cite{barone,tink,gross2016}. The experimental demonstration of the 
dc and ac Josephson effects in a dilute BEC was first proposed in Ref. \cite{giova00} and later 
effectively carried out
by utilizing a single JJ in relative motion with respect to the harmonic trap \cite{levy}. The authors
also discussed the experimental feasibility of an atom dc-SQUID, 
in particular its eventual application as a rotation
sensor based on the dependence of the critical current on the condensate
rotation rate \cite{levy}. On the other hand,
a similar dc-SQUID-type experimental setup with a pair of counterrotating weak links instead of the tunnel junctions
was implemented in 
Ref. \cite{jen14} to study the microscopic origin of the resistive flow appearing when the superfluid
current reaches the critical value. Previously, an important variant of these experiments had been
carried out for a single rotating weak link \cite{wright}, which is essentially analogous to a rf-SQUID. Such a configuration, which also has the potential to be utilized as a sensitive rotation sensor, was demonstrated 
to possess quantized hysteresis, in the first observation of such a phenomenon in
a superfluid BEC \cite{eckel}.

In this work, we analyze the dc to ac Josephson transition in an atom dc-SQUID similar
to that of the experimental setup in Ref. \cite{boshier13}. Our theoretical study makes use of
Gross-Pitaevskii (GP) simulations and a two-mode (TM) model, where only the ground state and
the first-excited stationary state of an asymmetric double-well toroidal condensate are taken into account 
to build
the dynamics \cite{cat14}. Note that such a model is not expected to describe accurately a far from equilibrium configuration like that occurring in the ac Josephson regime, nor an eventual transition from 
ac to dc. The TM model applied to an asymmetric double-well toroidal condensate was discussed in Ref. \cite{cat14},
where important effective interaction effects were incorporated in the model parameters 
\cite{cap13,je13}. On the other hand, the time dependence of the 
potential barriers (moving JJs), required to recreate the Josephson dc and ac regimes,
should give rise to a time-dependent boson field operator (in the Schr\"odinger representation)
that would yield extra terms in the Heisenberg and derived equations of motion (GP, TM).
This suggests that the simplest starting point to study this issue could be to build a
Bose-Hubbard (BH) Hamiltonian in a TM approximation. In fact, we derive extra terms due to the barrier motion
in such a Hamiltonian, as well as a hysteretic additional contribution to the Heisenberg
equation of motion of the boson field operator in the TM approximation.
The BH Hamiltonian in the limit of a macroscopic occupation of states yields a condensate energy
depending on an order parameter, represented by the phase difference and the particle imbalance, and
a control parameter, represented by the position and velocity of the JJs. The study of such an energy landscape 
allows to find the
location of the regions where the dc and ac Josephson effects are expected to occur, along with the corresponding 
critical points. 

We have considered two kinds of barrier motion. In the first one, 
an adiabatic initial acceleration leads to a final uniform velocity that ultimately yields the dc to ac transition. 
The second kind shares the same initial acceleration protocol up to reaching a maximum velocity, 
from where there is a symmetric decceleration leading to a final symmetric rest position of the barriers.
Barrier trajectories of this kind allow to observe the hysteretic effects that take place when the condensate
gets quite close to the dc-ac transition. Such effects are found to be 
similar to those occurring for faster barrier trajectories
able to drive the condensate from the dc to the ac regime and, 
in the subsequent stage of barrier decceleration, 
bring it back to the dc regime via the action of a resistive flow.
Similar hysteretic processes are common in superconducting JJs \cite{martinis90,castellano} and
have been studied within the frame of the resistively and capacitively shunted junction (RCSJ) model
\cite{stewart,*mccumber}. In our case, taking note of the close analogy between such a model 
and a TM model with damping \cite{marino},
we have analyzed the hysteresis loops that should take place as a result of 
the action of ac resistive flows.

This paper is organized as follows. In the next section, we specify
the technical details of the physical system considered in this study, along with the corresponding GP simulations.
Section \ref{secBH} treats the BH model and the TM equations of motion, starting with a review of the
case with barriers at rest in Sec. \ref{secRest}, and next by extending the treatment to the
case of moving barriers in Sec. \ref{secMoving}. 
The energy landscape, which allows to locate the regions belonging to 
the dc and ac regimes, is analyzed in Sec. \ref{secEnergy}, while the dc-ac transition and related hysteresis effects
for several adiabatic barrier trajectories are dealt with in Sec. \ref{secBarrierMotion}. Finally, a comparison
between the GP simulations and the TM model results is performed in Sec. \ref{secGPTM}, 
while some concluding remarks are gathered in Sec. \ref{secConclusion}.

\section{The System and GP simulations}
We describe in what follows the system utilized in our simulations and model applications.
All the trapping parameters and condensate details have been chosen to reproduce the experimental setting of Ref. \cite{boshier13}.
The trapping potential can be written as the sum of a part that depends only
on $x$ and $y$ and a part that is harmonic in the tightly bound direction $z$:
\begin{equation}
V_{\text{trap}}(x,y,z)=V(x,y)+\lambda^2z^2/2
\label{poten}
\end{equation}
being
\begin{equation}
V(x,y)=V_{\text{T}}(r)+V_{\text{B }}(x,y).
\end{equation}
The above potential consists of a superposition of
a toroidal term $V_{\text{T}}(r)$  ($r^2=x^2+y^2$)
and the radial barrier term   $V_{\text{B }}(x,y)$.
The toroidal potential was modeled through the following Laguerre-Gauss optical potential
\cite{lag}
\begin{equation}
V_{\text{T}}(r)=V_0\left[1-\left(\frac{r^2}{r_0^2}\right)\,\exp\left(1-\frac{r^2}{r_0^2}\right)\right],
\label{toro}
\end{equation}
where $V_0$ corresponds to the depth of the potential and $r_0$ the radial position
of its minimum.

The barriers were modeled as
\begin{equation}
V_{\text{B}}(x,y) =
V_b \,\, \sum_{k=1}^{2} \exp \left\{ -\frac{ [ y\cos\theta_k 
- x\sin\theta_k ]^2}  
{ \lambda_b^2}\right\} \Theta[y \sin\theta_k + x\cos\theta_k  ],
\label{barre}
\end{equation}
where $ \Theta $ denotes the Heaviside function with $\theta_1= \theta $ and 
$ \theta_2= \pi-\theta$.
 \begin{figure}
\includegraphics{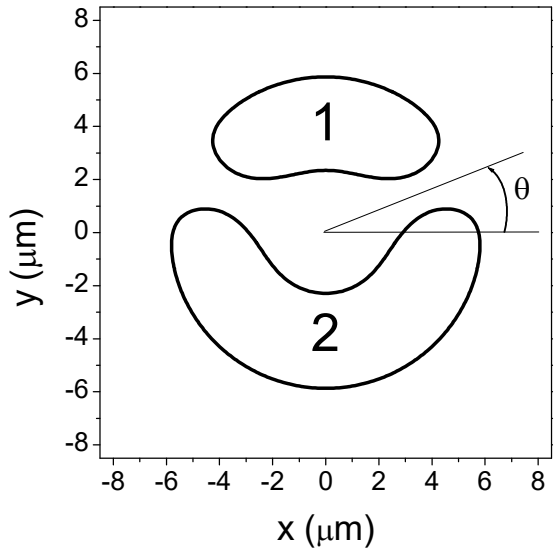}
\caption{Particle density isocontours for the ground state of $N$=3000. The barrier positions correspond to $\theta=\pi/8$.}
\label{figu1}
\end{figure}
 The parameter $\theta$ (see Fig. \ref{figu1}) may depend on time according to the
barriers movement, and
the following system parameters were utilized \cite{boshier13}:
  $V_0/k_B$ = 70 nK, $r_0$ = 4 $\mu$m,
$ V_b/k_B $ =  41.07 nK, $ \lambda_b = 1\, \mu$m and $N$=3000 atoms of $^{87}$Rb. 
We have used in our GP simulations scaled units referenced to the unit of length $L_0=1$ $\mu$m,
which yields energy and time units given by
$E_0/k_B=\hbar^2/(k_B mL_0^2)=5.5298$ nK and $T_0=\hbar/E_0=$1.3813 ms, respectively,
where $m$ denotes the mass of a condensate atom. 
We have assumed a high value of $\lambda$ in Eq. (\ref{poten}),
$(\lambda L_0)^2= 64\,E_0$, yielding
a quasi-bidimensional condensate and allowing a simplified
numerical treatment \cite{castin}.
So, the adimensionalized condensate order parameter is written as the product
of a Gaussian wave function along the $z$ coordinate,
$\sqrt{\frac{\lambda^{1/2}}{\pi^{1/2}}}\,\,e^{-\frac{\lambda z^2}{2}}$,
and a two-dimensional (2D) wave function $\psi(x,y,t)$ normalized to one,
for which the corresponding GP equation in scaled units reads
\begin{equation}
i \frac{\partial\psi}{\partial t}=
-\frac{1}{2}\left(\frac{\partial^2\psi}{\partial x^2}+\frac{\partial^2\psi}{\partial y^2}\right)+
V(x,y)\,\psi+gN\,|\psi|^2\psi,
\label{gp}
\end{equation}
where the effective $2D$ coupling constant $g=\sqrt{\frac{\lambda}{2\pi}}g_{3D}$ is written in terms
of the coupling constant between the atoms
$g_{3D}=\frac{4\pi\hbar^2a/m}{E_0L_0^3}=4\pi a/L_0$,
with $a= 98.98\, a_0 $ the  
$s$-wave scattering length of $^{87}$Rb and $a_0 $ the Bohr radius.
Such a GP equation was solved using the 
split-step Crank-Nicolson method \cite{adhikari} on a 2D spatial grid of 171$\times$171 points.

\section{BH model and TM equations of motion}\label{secBH}
\subsection{Barriers at rest}\label{secRest}

We begin by reviewing in this section the TM equations of motion for the toroidal 
asymmetric double-well condensate with barriers at rest derived in Ref. \cite{cat14}. However, 
in contrast to the previous treatment, we will start here from a BH model in order to facilitate
a generalization to the case of moving barriers in the next section.

The following BH Hamiltonian arises as usual from the many-body second-quantized Hamiltonian written in terms
of the TM approximation of the boson field operator $\hat \Psi (x,y)=\psi_1 (x,y) \hat a_1 + \psi_2 (x,y) \hat a_2$
 \cite{cat11}, where $\psi_k (x,y)$ denotes the real
wave function of a boson localized in the $k$-well with a
corresponding annihilation operator denoted by $\hat a_k$,
\begin{eqnarray}
\hat H_{BH} &=& \varepsilon_1 \hat a_1^{\dagger}\hat a_1 + \varepsilon_2 \hat a_2^{\dagger}\hat a_2
- K (\hat a_1^{\dagger}\hat a_2 + \hat a_2^{\dagger}\hat a_1) +
\frac{U_1}{2}\hat a_1^{\dagger}\hat a_1^{\dagger}\hat a_1\hat a_1 +
\frac{U_2}{2}\hat a_2^{\dagger}\hat a_2^{\dagger}\hat a_2\hat a_2 \nonumber \\
&-& F_{12}(\hat a_1^{\dagger}\hat a_1^{\dagger}\hat a_1\hat a_2 + 
\hat a_1^{\dagger}\hat a_2^{\dagger}\hat a_1\hat a_1)
-F_{21}(\hat a_2^{\dagger}\hat a_2^{\dagger}\hat a_1\hat a_2 + 
\hat a_1^{\dagger}\hat a_2^{\dagger}\hat a_2\hat a_2) \nonumber \\
&+& \frac{S}{2}(\hat a_1^{\dagger}\hat a_1^{\dagger}\hat a_2\hat a_2 + 
\hat a_2^{\dagger}\hat a_2^{\dagger}\hat a_1\hat a_1 + 4 \hat a_1^{\dagger}\hat a_2^{\dagger}\hat a_1\hat a_2),
\label{BH}
\end{eqnarray}
where
\begin{equation}
\varepsilon_k = \int d^2r\,\,  \psi_k(x,y) \left[
-\frac{ \hbar^2 }{2 m}{\bf \nabla}^2 +
V(x,y)\right]  \psi_k(x,y),
\label{epsR}
\end{equation}

\begin{equation}
K= -\int d^2r\,\, \psi_{1}(x,y) \left[
-\frac{ \hbar^2 }{2 m}{\bf \nabla}^2  +
V(x,y)\right]  \psi_{2}(x,y),
\label{jota0}
\end{equation}

\begin{equation}
U_k= g \int d^2r\,\,  \psi_k^4(x,y),
\label{U0R}
\end{equation}

\begin{equation}
F_{jk}=   -g\int d^2r\,\,  \psi_{j}^3(x,y)
 \psi_{k}(x,y),
\label{FRL}
\end{equation}

\begin{equation}
S= g  \int d^2r\,\,   \psi_{1}^2(x,y) \,  \psi_{2}^2(x,y) .
\label{ijota}
\end{equation}
Here it is important to remark that such localized states, characterized by the wave functions $\psi_{k}(x,y)$ and the corresponding
operators  $\hat a_k$ and $\hat a_k^{\dagger}$, may actually depend on the number of particles at each well. 
Particularly,
as a most significant effect of the repulsive interparticle interaction, there is a broadening 
of the wave functions $\psi_{k}$ with increased occupation numbers \cite{cat11,duttar}. 
However, since the occupation number variations for our time evolutions in the dc Josephson regime, 
including the transition to the ac regime, actually keep small enough,
we will disregard, in principle, 
such a dependence in the Hamiltonian (\ref{BH}). Nevertheless, in the following we shall see 
that the above interaction effects may be taken into account at the level of
the TM equations of motion, in order to get a substantial improvement in the agreement
with the GP simulation results.

The Hamiltonian (\ref{BH}) rules the condensate dynamics according to the Heisenberg equations,
\begin{equation}
\frac{d\hat a_k}{dt}=\frac{i}{\hbar}[\hat H_{BH},\hat a_k],
\label{heis}
\end{equation}
and assuming a macroscopic occupation of states, one may replace the annihilation operators by complex $c$-numbers,
\begin{equation}
\hat a_k\rightarrow \sqrt{N_k}\, \exp(i\phi_k), 
\label{repl}
\end{equation}
where $\phi_k$  and $N_k$ represent the phase and particle number in
the $k$-well, respectively. By performing such a replacement in (\ref{heis}) 
one may obtain the following equations of motion of
the TM model, where the time evolution of the condensate is described through the particle imbalance
$Z=(N_2-N_1)/N$ and the phase difference between both wells $\phi=\phi_1-\phi_2$ (see Fig. \ref{figu1}),
\begin{eqnarray}
 \hbar  \dot{Z} & =& -\frac{\Delta E}{(1-Z_0^2)^{3/2}} 
\sqrt{1-Z^2}\,(1-2Z_0^2+Z_0Z)\sin\phi   +   N S \,  (1 - Z^2) \sin (2 \phi) \label{zpun}\\
\hbar  \dot{\phi}  & =& (Z-Z_0)\left[  N(U_1+U_2)/2 
+ \frac{\Delta E}{(1-Z_0^2)^{3/2}} \frac{(1+2Z_0Z)}{\sqrt{1-Z^2}}\cos\phi -2NS\right] \nonumber\\  
&+&  N S [Z_0- Z \cos (2 \phi)] \,, \label{fipun}
%\nonumber\\ 
%&\simeq& (Z-Z_0)N(U_1+U_2)/2 \,, \label{fipun}
\end{eqnarray}
where $\Delta E=N(F_{21}-F_{12})(1-Z_0^2)^{3/2}/Z_0$ 
denotes the energy-per-particle splitting between both stationary solutions of the above equations of motion,
namely the ground state with $\phi=0$ and the first-excited state with $\phi=\pm\pi$, both states having the same
imbalance $Z=Z_0$, whose value depends on the barrier angle $\theta$. 
Such states, which correspond to the stationary solutions of the GP equation
(\ref{gp}) yielding the lowest condensate energies, 
can be expressed as linear combinations of the localized states,
\begin{equation}
\psi_\pm = \pm\sqrt{\frac{1-Z_0}{2}}\psi_1+\sqrt{\frac{1+Z_0}{2}}\psi_2,
\end{equation}
where $\psi_+$ ($\psi_-$) denotes the wave function of the ground (first-excited) single-particle state.
However, such wave functions  are not mutually orthogonal, except for a symmetric configuration ($\theta=0$ in
Fig. \ref{figu1}) \cite{cat14}. In fact, for a whole population $N$ in the ground (first-excited) single-particle state, 
there would be $NZ_0^2$ atoms in the first-excited (ground) single-particle state.

The hopping contributions proportional to the energy gap $\Delta E$ and the parameter $S$ 
in (\ref{fipun}),
turn out to be quite negligible with  respect to the term proportional to the average on-site interaction energy 
$(U_1+U_2)/2$, so we may adopt such an approximation in Eq. (\ref{fipun}). 
In addition, we have seen in Ref. \cite{cat14}
that the agreement between the TM time evolution results and the GP simulations turns out to be
substantially improved by
replacing such an average on-site interaction energy by the following expression,
\begin{equation}
U+BZ \label{Umo}
\end{equation}
with
\begin{eqnarray}
U & =  &  \frac{1}{2}   [ (1- \alpha_{1} ) U_{1} +  (1- \alpha_{2}  ) U_{2})]   \nonumber\\
&& \label{bmo}\\
B  & =  & \frac{1}{2}   (  \alpha_{1}  U_{1}  -  \alpha_{2}   U_{2} )   \, , \nonumber
\end{eqnarray}
where the parameters $\alpha_j$ arise from the deformation that suffer the condensate densities at both wells
due to the departure of the particle imbalance $Z$ from the stationary value $Z_0$ during the time evolution. 
In fact, denoting by
$\rho_{j}^{(\Delta N_{j})}$ the probability density of the localized state at site $j$ with
$N_j^0+\Delta N_j$ particles,
where $N_j^0$ denotes the population of the $j$-well corresponding to the imbalance $Z_0$, 
the parameter $\alpha_j$ may be obtained from the 
following modified on-site interaction energy parameter \cite{cap13,je13,cat14}
\begin{eqnarray}
 U^{(\Delta N_{j}) }_{j} &  = & g \int d^2r\,\,   \rho_j^{(0)}(x,y) \,   \rho_{j}^{(\Delta N_{j})}(x,y) \nonumber\\
&  \simeq & (1-2 \alpha_j  \frac{\Delta N_j }{N})  U_{j},
\label{urdelta}
\end{eqnarray}
where $\rho_j^{(0)}=\psi_j^2$  and the second line corresponds to the first-order approximation on $\Delta N_j /N$,
being $  2 \Delta N_{2}/N = -  2 \Delta N_{1} /N= Z -Z_0$.
To extract, in practice, the parameter $\alpha_j$, one should numerically 
evaluate the above integral with a probability density
$\rho_{j}^{(\Delta N_{j})}(x,y)$ obtained from the stationary states  of a condensate
with $N_j^0+\Delta N_j$ particles on the site $j$ and
 a total number of particles that differs from $N$ \cite{cap13,je13,cat14}.
The idea behind such calculations is 
that a nonequilibrium state should be well aproximated by localized on-site states
corresponding to the instantaneous populations at each well. Thus, the replacement 
in Eq. (\ref{fipun})
of the average on-site interaction energy $(U_1+U_2)/2$  by the imbalance-dependent parameter 
$U+BZ$, serves to quantitatively taking into account the above density variations.
It is important to remark that the contributions arising from the parameters $\alpha_j$ in 
(\ref{bmo}) turn out to be far from negligible, despite of being originated at the first-order approximation 
(\ref{urdelta}). In fact, such effective interaction corrections were found to substantially improve the agreement
with the time-dependent GP simulation results  \cite{cat14}. 
Finally, we may rewrite the equation of motion (\ref{fipun}) as,
\begin{eqnarray}
\hbar  \dot{\phi} &  = & (Z-Z_0)  N( U  +  B Z) \nonumber \\
&=&\mu_2-\mu_1, \label{fip}
\end{eqnarray}
where $\mu_j$ denotes the chemical potential of the condensate at the site $j$. 
The above equivalence between
the time derivative of the phase difference at both sides of a Josephson junction
and the corresponding chemical potential difference has been shown to possess a wide 
range of validity \cite{satop,levy}, and it is analogous to the voltage-phase relation of the Josephson effect in superconductors 
\cite{barone,tink,gross2016}.

The condensate energy in the TM approximation may be obtained from the BH Hamiltonian 
(\ref{BH}) under the replacement  (\ref{repl}), yielding
\begin{eqnarray}
E_{TM}&=& N\frac{(\varepsilon_1+\varepsilon_2)}{2}+N^2\frac{(U_1+U_2)}{4}\left(\frac{1+Z^2}{2}-Z_0Z\right)
\nonumber\\
&-& \frac{\sqrt{1-Z^2}}{2(1-Z_0^2)^{3/2}}N\Delta E (1-2Z_0^2+ZZ_0)\cos \phi\nonumber\\
&+& \frac{N^2S}{2}\left[(1-Z^2)\left(1+\frac{1}{2}\cos 2\phi\right)+3Z_0Z\right].
\end{eqnarray}
However, it will be more convenient to measure the energy relative to that of the ground-state ($Z=Z_0$, $\phi=0$),
which yields
\begin{eqnarray}
\Delta E_{TM}\equiv E_{TM}-E_{TM}^{GS}&=& N^2\frac{(U_1+U_2)}{8}(Z-Z_0)^2
\nonumber\\
&+& \frac{N}{2}\Delta E\left[1-
\frac{\sqrt{1-Z^2}}{(1-Z_0^2)^{3/2}} (1-2Z_0^2+ZZ_0)\cos \phi\right]\nonumber\\
&+& \frac{N^2S}{2}\left\{(1-Z^2)\left(1+\frac{1}{2}\cos 2\phi\right)+\frac{3}{2}[2Z_0Z-(1+Z_0^2)]\right\},
\label{ener}
\end{eqnarray}
where it is worthwhile noticing that the first-excited single-particle state ($Z=Z_0$, $\phi=\pm\pi$) yields
the correct energy gap $\Delta E_{TM}=N\Delta E$ in (\ref{ener}). 
Note also that the equations of motion (\ref{zpun})-(\ref{fipun}) can be 
written in the Hamiltonian form
\begin{equation}
\hbar \dot N_2 = -\frac{\partial {\cal H}}{\partial\phi}\,\, ; \,\, \hbar \dot\phi = \frac{\partial {\cal H}}{\partial N_2},
\label{form}
\end{equation}
being ${\cal H}=\Delta E_{TM}$ the Hamiltonian and  ($N_2$,$\phi$) the canonically conjugate variables.
Here it is instructive to approximate the expression (\ref{ener}) for $Z\simeq Z_0$, along with 
neglecting the term proportional to the second-order hopping parameter $S$.
Thus, taking into account Eq. (\ref{fipun}) without the tunneling  contributions,
we obtain the following Hamiltonian for a ``phase particle" of coordinate $\phi$
\begin{equation}
{\cal H}\simeq \frac{\hbar^2\dot\phi^2}{2(U_1+U_2)}+\frac{N\Delta E}{2}(1-\cos\phi),
\label{phasepart}
\end{equation}
where the first and second term should be interpreted as the kinetic and potential energy,
respectively. Note that the minimum and maximum of the potential energy at $\phi=0$ and $\phi=\pm\pi$, respectively,
correspond to the above stationary states, as expected. A similar expression
is found for the Hamiltonian of a superconducting Josephson junction in the low damping limit, 
which may be quantized through a straightforward procedure to investigate the quantum behavior
of the phase difference \cite{clarke,blackburn}. 
However, such quantum effects turn out to be quite negligible in our case,
since the frequency $\omega_p$
of the low amplitude (plasma) oscillations around the potential minimum, yielding the level spacing 
$\hbar\omega_p=\sqrt{N\Delta E(U_1+U_2)/2}$, turns out to be much
smaller than the depth of the potential well $N\Delta E$, a result 
which agrees with the classical picture represented by the Hamilton equations (\ref{form}).

\subsection{Moving barriers}\label{secMoving}

\begin{figure}
\includegraphics{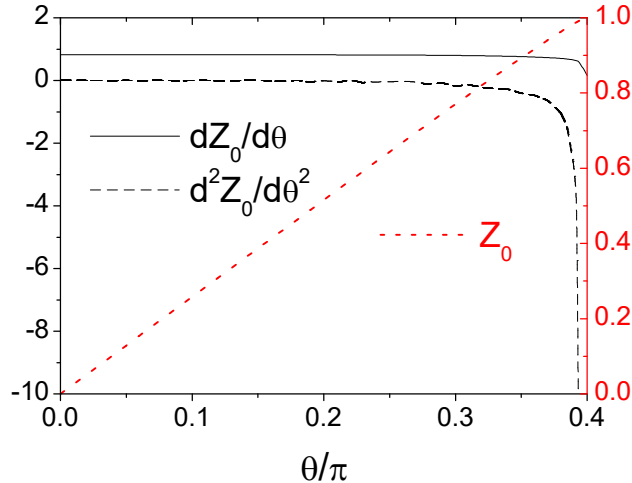}
\caption{The stationary imbalance $Z_0$ and its first two derivatives versus the barrier angle $\theta$. The right scale corresponds to $Z_0$,
while the left scale corresponds to its derivatives.}
\label{figu2}
\end{figure}

If the barriers are set in motion, there are two effects that could
modify the Heisenberg equation for the boson field operator
$\hat \Psi$.
The first effect stems from the additional contribution to the many-body Hamiltonian
arising from the induced particle flux across the moving barriers. A second effect should
arise from the time dependence of the boson field operator in the Schr\"odinger representation
due to the barrier motion.
Let us first discuss the additional contribution to the energy.
In fact, as barriers are displaced, the change in the particle number of each localized state causes a local
energy change at a rate given by the corresponding chemical potential $\mu_j=\partial E_j/\partial N_j$. So, taking 
into account such contributions, the additional energy may be written,
\[\int_{N_1^{\rm ref}}^{N_1}\mu_1\, dN'_1 + \int_{N_2^{\rm ref}}^{N_2}\mu_2\, dN'_2=
\int_{N_2^{\rm ref}}^{N_2}(\mu_2-\mu_1)\, dN'_2,\]
where the values $N_j^{\rm ref}$ denote reference (initial) particle numbers. Note that have dropped the reference energies and
have taken into account the constraint $N'_1+N'_2=N$ in the last equality.
To proceed with the calculation, we may change variables in
the last integral to the time domain and use (\ref{fip}) to obtain
\[\hbar\int_0^t  \frac{d\phi}{dt'}\,I^0(t')dt',\]
where $I^0\equiv\dot N_2^0=(N/2)(\partial Z_0/\partial\theta)\,\dot\theta$ 
will be called the bias current, as it is kinematically dependent of the barrier motion, 
to distinguish from the actual particle current $I\equiv\dot N_2$.
Next, we integrate the above result by parts in order to split it into an `instantaneous' contribution,
\begin{equation}
\hbar I^0(t)\phi(t),
\label{eadic}
\end{equation}
where we have dropped its initial value,
and a hysteretic contribution,
\begin{equation}
-\hbar\int_0^t  \dot I^0(t')\phi(t')dt',
\label{eadich}
\end{equation}
which depends on the previous history of the system. Note that both contributions will be proportional to the time
derivatives of the stationary imbalance $Z_0(\theta(t))$,
\begin{eqnarray}
2\frac{I^0}{N} & = & \dot Z_0  = \frac{\partial Z_0}{\partial\theta}\,\dot\theta, \label{dot} \\
2\frac{\dot I^0}{N} & = & \ddot Z_0  =  \frac{\partial^2 Z_0}{\partial\theta^2}\,\dot\theta^2+
\frac{\partial Z_0}{\partial\theta}\,\ddot\theta.\label{ddot}
\end{eqnarray}
We depict in Fig. \ref{figu2} such an imbalance as a function of the barrier angle $\theta$, along with its first two derivatives.
There we may observe that $Z_0$ presents a rather linear behavior, yielding a constant value for the first derivative and a negligible value
for the second, except for barrier angles approaching $0.4 \pi$, where the barriers begin to overlap each other. Particularly, the second
derivative becomes clearly nonnegligible for such configurations. For instance, in the simplest situation of a
barrier motion with a constant velocity, 
there would be an initial jump in velocity from the previous configuration of barriers at rest, 
which would yield a Dirac delta acceleration 
$\ddot\theta\sim\delta(t)$ in (\ref{ddot}) and in the integrand of the expression (\ref{eadich}). However, this contribution to the energy
should be irrelevant, since it could be treated as a constant term. 
On the other hand, the first term of
(\ref{ddot}) could become nonnegligible for barrier angles approaching $0.4 \pi$, and this could 
eventually activate the hysteretic contribution (\ref{eadich}) to the condensate energy.

The additional energy (\ref{eadic}) has previously been considered 
in the context of an analogy with the RCSJ model,
widely applied for superconducting Josephson junctions
 \cite{barone,tink,gross2016,clarke,blackburn,giova00,levy,scazza}. 
However, its corresponding counterpart in the quantum many-body Hamiltonian, along with an
eventual effect on the equations of motion, e.g., on the time-dependent GP equation,
seem to be so far undiscussed. Here it is important to recall 
that experimental results on similar configurations of
 ring-shaped condensates with moving barriers, have shown
a good agreement with the corresponding simulation results 
arising from the GP equation without any correction, 
apart from the time dependence of the trapping potential due to the moving barriers. Such experiments 
were conducted for different kinds of barrier motion, namely
for adiabatic accelerations up to the final constant velocity \cite{boshier13}, as well as
for a sudden set into motion of the barriers at a constant speed \cite{jen14}.

Now, turning back to the theoretical arena, we should seek for
 an additional contribution to the Hamiltonian (\ref{BH}), which in the limit (\ref{repl}) 
of a macroscopic occupation, would yield the expression (\ref{eadic}). 
A complete solution of this problem should be closely related to the long-standing 
and quite delicate issue of obtaining an acceptable quantum description 
of the phase by means of a phase operator \cite{barnett}. Rather than pursuing
such an ambitious goal, we will content ourselves
with restricting our analysis to a limit of high populated
localized states.
In fact, let us consider the following operator,
\begin{equation}
\frac{i}{2}\left[\ln(\hat a_1^\dagger\hat a_2)-\ln(\hat a_2^\dagger\hat a_1)\right],
\label{logarithm}
\end{equation}
which would yield in the limit (\ref{repl}) just the phase difference $\phi$. However, in defining the above expression
one should take into account that the existence of the logarithm
of a given operator must require that it should be invertible \cite{higham}. In particular, the creation and annihilation operators
only partially fulfill such a requirement, since they turn out to be, 
similarly to the Susskind-Glogower phase operators \cite{suss},
one-sided unitary, namely \cite{saxena}
\begin{eqnarray}
\hat a_k\hat a_k^{-1}&=&\hat a_k^{\dagger-1}\hat a_k^{\dagger}={\bf I}\\
\hat a_k^{-1}\hat a_k&=&\hat a_k^{\dagger}\hat a_k^{\dagger-1}={\bf I}-|0\rangle\langle 0|,
\label{onesided}
\end{eqnarray}
where $\hat a_k^{-1}$ and $\hat a_k^{\dagger-1}$ are one-sided inverse operators, ${\bf I}$ is the identity operator and
$|0\rangle\langle 0|$ denotes the projection operator on the vacuum of the $k$-well. 
Such a projection, however,
should be quite irrelevant in our case, given the limitations we have assumed 
on the subspace where the Hamiltonian (\ref{BH}) should be supposed to be acting upon, i.e.,
limited only to high occupation configurations. So, we will disregard in what follows 
the last term in (\ref{onesided}), which in turn leads to a well-defined operator (\ref{logarithm}) that,
combined with expressions (\ref{eadic}) and (\ref{eadich}),
suggests that the following additional terms to the BH Hamiltonian (\ref{BH}) should be
taken into account:
\begin{eqnarray}
\hat H_1 &=&\frac{i\hbar I^0(t)}{2}\{\ln[\hat a_1^\dagger(t)\hat a_2(t)]-\ln[\hat a_2^\dagger(t)\hat a_1(t)]\} \\
\hat H_2 &=&-\frac{i\hbar}{2}\int_0^t \dot I^0(t')
\{\ln[\hat a_1^\dagger(t')\hat a_2(t')]-\ln[\hat a_2^\dagger(t')\hat a_1(t')]\}dt'.
\label{h_2}
\end{eqnarray}
We may now rewrite the Heisenberg equation (\ref{heis}) in the case of moving barriers as,
\begin{equation}
\frac{d\hat a_k}{dt}=\frac{i}{\hbar}[\hat H_{BH}+\hat H_1+\hat H_2,\hat a_k]+
\frac{\partial\hat a_k}{\partial t},
\label{heisen}
\end{equation}
where the partial derivative denotes a time derivative of the Schr\"odinger operator, which next is time-evolved 
to reach the Heisenberg picture. To evaluate the commutator $[\hat H_1,\hat a_k]$ in (\ref{heisen}), 
we make use of the following result \cite{huele},
\[[\hat a_k,\ln\hat a_k^\dagger]=\frac{\partial(\ln\hat a_k^\dagger)}{\partial\hat a_k^\dagger}=\hat a_k^{\dagger-1},\]
which yields,
\begin{equation}
\frac{i}{\hbar}[\hat H_1,\hat a_k]=-\frac{\dot N_k^0}{2}\,\hat a_k^{\dagger-1}.
\label{conmut}
\end{equation}
On the other hand, to evaluate the partial derivative in (\ref{heisen}) it is convenient, at a first stage, to do it in the limit 
(\ref{repl}). Then, we have in the Schr\"odinger picture, $(\hat a_k)_{\rm S}\rightarrow\sqrt{N_k^0(t)}\exp(i\phi_k^0)$, where only
the population of the localized states will be time-dependent as barriers move. Therefore, the time derivative turns out to be,
\begin{equation}
\left(\frac{\partial\hat a_k}{\partial t}\right)_S\rightarrow\frac{\dot N_k^0}{2\sqrt{N_k^0}\exp(-i\phi_k^0)}\leftarrow
\frac{\dot N_k^0}{2}(\hat a_k^{\dagger-1})_S,
\label{sch}
\end{equation}
which also turns out to arise from the limit (\ref{repl}) of the above right-hand side expression. Finally,
turning to the Heisenberg representation in (\ref{sch}), we may conclude that the term of
the partial time derivative in (\ref{heisen}) and the term (\ref{conmut}) cancel each other, yielding the following
Heisenberg equation of motion,
\begin{equation}
\frac{d\hat a_k}{dt}=\frac{i}{\hbar}[\hat H_{BH}+\hat H_2,\hat a_k].
\label{heisenberg}
\end{equation}
Here an eventual calculation of the commutator $[\hat H_2,\hat a_k]$ in the above expression 
appears as a quite difficult
task, as it involves the evaluation of commutators of Heisenberg operators at different times.
However, according to (\ref{h_2}) and (\ref{ddot}), such a hysteretic contribution to the equation
of motion should be negligible, except for accelerated or nearly overlapping barriers.
Moreover, no evidence of a discrepancy between the experimental results for adiabatically accelerated barriers 
and the corresponding GP simulations has been reported so far \cite{levy,boshier13}. Thus, we will disregard in
what follows any hysteretic contribution stemming from (\ref{eadich}) or (\ref{h_2}) to the energy and the equations of motion,
both for GP simulations and the TM model. Finally to summarize, we will assume 
that the Heisenberg equation of motion remains 
formally equivalent to the original expression (\ref{heis}), except for the parameters of the BH
Hamiltonian (\ref{BH}) that now 
become time-dependent as barriers move. The same occurs with the TM equations of motion
(\ref{zpun}) and (\ref{fipun}), along with the effective interaction correction (\ref{fip}). On the other hand, given that
the energy represented by ${\cal H}$ in the Hamiltonian formalism (\ref{form}) acquires  for moving barriers
the additional term (\ref{eadic}),
we may generalize Eqs. (\ref{form}) as,
\begin{equation}
\hbar \Delta \dot N_2 = -\frac{\partial {\cal H}}{\partial\phi}\,\, ; \,\, \hbar \dot\phi = \frac{\partial {\cal H}}{\partial\Delta N_2},
\label{formh}
\end{equation}
where now the Hamiltonian is given by ${\cal H}=\Delta E_{TM}+\hbar I^0\phi$,
and ($\Delta N_2$,$\phi$) represent the canonically conjugate variables,
with $\Delta N_2=N_2-N_2^0$. In addition, the potential energy of the ``phase particle"  in (\ref{phasepart}) becomes
for moving barriers a ``tilted washboard" potential 
\cite{barone,tink,gross2016,clarke,blackburn,giova00,levy,scazza} given by,
\begin{equation}
\hbar I_c\left(1-\cos\phi+\frac{I^0}{I_c}\phi\right),
\label{tilted}
\end{equation}
with
\begin{equation}
I_c=\frac{N\Delta E}{2\hbar}
\label{crit}
\end{equation}
the so-called Josephson critical current.  Here it is worthwhile noticing that all the currents we have 
defined so far, i.e.,
the particle current $I=\dot N_2$, the bias current $I^0=\dot N_2^0$ and the critical current $I_c$, actually
are twice the corresponding current across each JJ.
The average slope of the ``washboard" in (\ref{tilted})
is proportional to the quotient
$I^0/I_c$ and the local extrema of such a potential arise from the Josephson current-phase relation \cite{barone,tink,gross2016,golubov}
\begin{equation}
I^0=-I_c\sin\phi.
\label{sinu}
\end{equation}
Particularly, the
stationary states for barriers at rest ($I^0=0$) located at $\phi=0$ (minimum) and $\phi=\pm\pi$ (maximum) turn out to be
displaced for moving barriers to the following locations,
\begin{equation}
\phi_m=-\sin^{-1}\left(\frac{I^0}{I_c}\right),
\label{fi_m}
\end{equation}
and
\begin{equation}
\phi_M=\sin^{-1}\left(\frac{I^0}{I_c}\right)-\frac{I^0}{|I^0|}\pi,
\label{fi_M}
\end{equation}
where $\phi_m$ and $\phi_M$ denote the phase differences of the minimum and maximum of the washboard potential
within the interval $-\pi<\phi<\pi$, respectively.
The above results can equivalently be expressed in terms of the barrier speed by means of Eq. (\ref{dot}),
\begin{equation}
\frac{I^0}{I_c}=\frac{f}{f_c},
\label{ffc}
\end{equation}
where $f=\dot\theta/(2\pi)$ denotes the barrier rotation frequency and
\begin{equation}
f_c=\frac{\Delta E}{h(\partial Z_0/\partial\theta)},
\label{fc}
\end{equation}
represents the critical rotation frequency corresponding to the Josephson critical current (\ref{crit}). 
We may see from (\ref{fi_m}) and (\ref{ffc}) that the phase particle can remain ``trapped" around the potential minimum
provided the bias current does not exceed the critical value $I_c$, or equivalently, the barrier speed remains
below the critical frequency $f_c$. Such a dynamics, characterized by a bounded phase difference,
vanishing average values for $\dot\phi$ and $\mu_2-\mu_1$ (\ref{fip}), and a
particle current that matches the bias current, corresponds to the so-called dc Josephson regime (analogous to the 
superconducting zero-voltage state \cite{barone,tink,gross2016}).
Otherwise, for a bias current or a barrier speed above the critical values, the local extrema of the washboard potential disappear and
the particle will ``fall down" indefinitely, yielding a dynamics  of running phase (nonvanishing values for $\langle\dot\phi\rangle$ and
$\langle\mu_2-\mu_1\rangle$) and an alternating particle current. 
Such characteristics define the so-called ac Josephson regime, which is analogous to the 
superconducting nonzero-voltage state \cite{barone,tink,gross2016}. This regime can also be accessed 
from below the critical values, provided the energy of the particle ${\cal H}$ exceeds 
the maximum value of the potential to escape over the top of the barrier and propagate down
the washboard. We disregard here any additional way of escaping, such as, by thermal activation at finite temperatures, or
by a macroscopic quantum tunneling process \cite{clarke,blackburn}, given 
the small relative value of the plasma frequency $\omega_p$ in our case.

\subsubsection{Energy landscape}\label{secEnergy}
\begin{figure}
\includegraphics{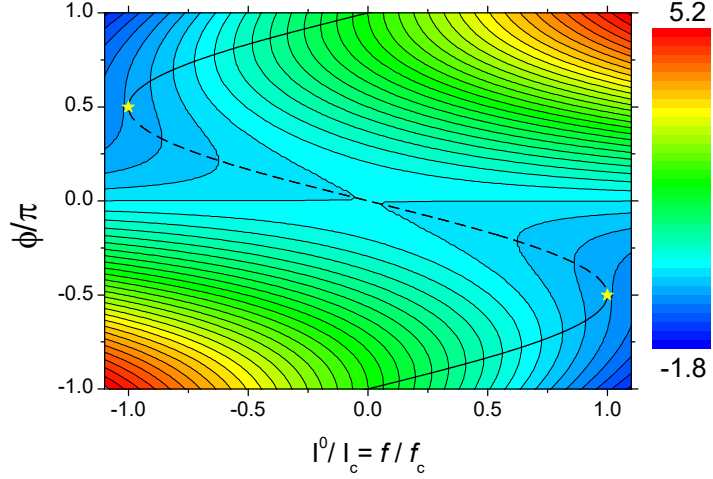}
\caption{Energy landscape for $Z=Z_0$. The color scale corresponds to the adimensionalized energy 
$(\Delta E_{TM}+\hbar I^0\phi)/\hbar I_c$, while the dashed (solid) line locates its minimum (saddle) for each value
of the bias current or barrier rotation frequency. 
The yellow stars indicate the critical points where minimum and saddle
coalesce. }
\label{figu3p}
\end{figure}

\begin{figure}
\includegraphics{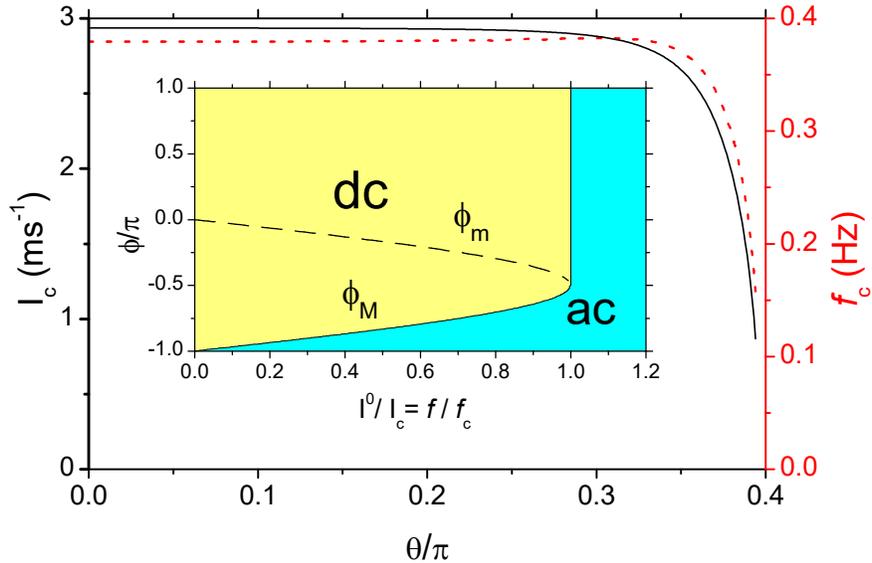}
\caption{Josephson critical current $I_c$ (Eq. (\ref{crit}), solid line) and critical frequency $f_c$ 
(Eq. (\ref{fc}), dotted line) versus the barrier position $\theta$. Inset: the dc (ac) Josephson regime 
corresponds to bias currents or barrier frequencies below (above) the critical values. Particularly,
the ac domain extends below the critical values for phase differences beyond that of the energy saddle 
$\phi_M$ (Eq. (\ref{fi_M})). On the other hand, the phase difference $\phi_m$ (Eq. (\ref{fi_m})) of the energy minimum
is depicted as a dashed line. }
\label{figu3}
\end{figure}

Although the above heuristic model of a fictitious particle moving in the washboard potential 
may be useful to understand the dynamics of the phase 
within the different Josephson regimes, such a model is based on several approximations that eventually could lead to
inaccuracies with respect to the GP simulation results. For instance, the term corresponding to the kinetic energy in 
(\ref{phasepart}) was derived from the equation (\ref{fipun}), instead of the more accurate expression (\ref{fip}),
in addition, the parameters of the kinetic energy and the washboard potential (\ref{tilted}) may depend
on the barrier position (see, e.g., Eqs. (\ref{dot}), (\ref{crit}) and Fig. \ref{figu3}), leading to a nonconserved energy
that could eventually invalidate such a model.
So, to avoid such shortcomings, we may utilize
a more formal treatment which consists in studying 
the condensate energy landscape as a functional of the order parameter \cite{mueller}.
This in our case amounts to
study the energy ${\cal H}=\Delta E_{TM}+\hbar I^0\phi$ as a function of the phase difference and the particle
imbalance. Previously we have seen that
the condensate energy 
with barriers at rest (\ref{ener}) presents a minimum for the ground-state
$\phi=0$ and a maximum (saddle) for the excited state $\phi=\pm\pi$, both for $Z=Z_0$.
Now, by canceling the partial derivatives in (\ref{formh}),
we may easily find such extrema for a general case of moving barriers. Thus, we encounter that 
the energy should present local extrema for $Z=Z_0=2N_2^0/N-1$ and phase differences fulfilling the following equation,
\begin{equation}
I^0=-I_c\sin\phi+(2NS N_2^0/\hbar)(1-N_2^0/N)\sin2\phi ,
\end{equation}
which, taking into account that the second harmonic contribution of the last term \cite{golubov,scazza}
can be safely neglected for our condensate, becomes equivalent to the previously considered simple sinusoidal 
current-phase relation (\ref{sinu}).
Therefore, according to the sign of the second derivatives of ${\cal H}(\Delta N_2,\phi)$,
there must be a minimum of the energy at $Z=Z_0$ and $\phi=\phi_m$ given by Eq. (\ref{fi_m}),
and a saddle (minimum for $Z$ and maximum for $\phi$) at $Z=Z_0$ and $\phi=\phi_M$ given by Eq. (\ref{fi_M}).
Particularly, the $Z$-dependence of the energy turns out to be 
quite simple since it is dominated by the harmonic term $\sim (Z-Z_0)^2$ in (\ref{ener}).
So, we can obtain a full picture of the energy landscape by setting $Z=Z_0$ and analyzing 
the remaining dependence on the phase
difference and the bias current, as depicted in Fig. \ref{figu3p}. A typical system trajectory in the dc regime 
will reside in
the vicinity of the dashed line, which corresponds to the phase difference at the 
energy minimum for each value of the bias current. 
On the other hand, the ac regime will be attained when exceeding the critical value of the bias current (or the barrier speed), as well as if the system acquires enough energy to overcome the saddle (solid) lines.
We note that Fig. \ref{figu3p} can also be regarded as an energy landscape depending on
the order parameter $\phi$ and the control parameter \cite{mueller} $I^0/I_c=f/f_c$ (see Eqs. (\ref{ffc})
and (\ref{fc})), where the latter
actually embodies two independent control parameters, the barrier position and its velocity. 
However, it will be seen in Fig. \ref{figu3} that
the dependence on the barrier position of $I_c$ and $f_c$ turns out to be very weak for $\theta\lesssim 0.3\pi$, 
so the problem within such an interval becomes reduced to the simplest one of a single
control parameter (bias current $I^0$, or rotation frequency $f$).
In fact, we display in Fig. \ref{figu3} the Josephson critical current and the corresponding critical frequency as 
functions of the barrier position, 
where we may observe that they remain practically constant until the barriers begin to overlap each other, 
a process which eventually may cause them to drop toward vanishing values. 
%As will be seen in Sec. \ref{secGPTM}, the values of the above critical parameters turn out to be slightly underestimated with respect to that arising from the GP simulations by about a 0.5 \%.
On the other hand, the inset of
such a figure shows the domains of the dc and ac Josephson regimes on a phase versus bias current (or 
rotation frequency)
diagram. We note that according to Fig. \ref{figu3p} a straightforward 
extension of such a diagram to negative abscissas could be easily depicted. 

\subsubsection{Adiabatic barrier motion: dc-ac transition and hysteretic phenomena}\label{secBarrierMotion}
Starting from an initial condition with the barriers at rest and the condensate in the ground state, any barrier motion will
trigger an oscillation of the order parameter, in particular, the particle current and the phase difference, 
which for small amplitudes will be ruled by the plasma frequency $\omega_p$. However, since we are interested in
the dc to ac transition driven by a current bias, and being the departure of the actual current from
the bias current one of the main signatures of such a transition, one should try to suppress, or at least minimize,
such oscillations, as they could certainly interfere with our observations. 
To this aim, we have assumed the following barrier trajectory that preserves
the continuity of the acceleration $\ddot\theta$ along the whole path,
\begin{equation}
\theta(t)=-\frac{\theta(0)}{\pi}[\sin(\omega t+\pi)+\omega t-\pi],\,\,\,\,\,\omega=-\frac{\pi^2 f_{\rm max}}{\theta(0)},
\label{traj}
\end{equation}
where $f_{\rm max}$ denotes the maximum value of the rotation frequency $f=\dot\theta/(2\pi)$ that
may be regarded as a control parameter of
the approach to the dc-ac transition. For instance, for $f_{\rm max}>f_c$ we have that
such a transition will certainly be reached during the trajectory. 
On the other hand, in order to reduce the angular
frequency $\omega$ in (\ref{traj}), increasing the adiabaticity, 
we will assume a large value of the initial angle $|\theta(0)|$ compatible with
nonoverlapping barriers, namely
$\theta(0)=-0.394\pi$. 
\begin{figure}
\includegraphics{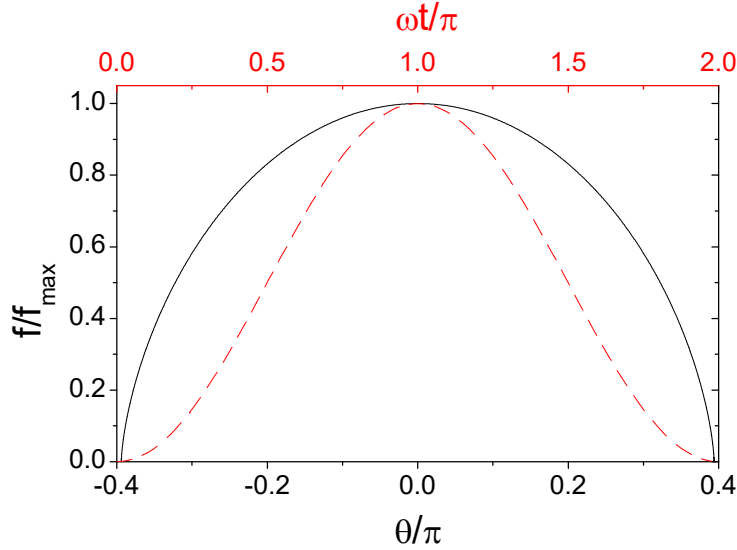}
\caption{Adiabatic barrier motion. Time evolution of the barrier rotation frequency $f=\dot\theta/(2\pi)$
 arising from Eq. (\ref{traj})
in units of its maximum value $f_{\rm max}=f(\theta=0)$ (dashed line),
and the same quantity 
versus the barrier position $\theta$ (solid line).}
\label{figu4}
\end{figure}
We depict in 
Fig. \ref{figu4} the rotation frequency $f=\dot\theta/(2\pi)$ versus time,
as well as its dependence on the barrier position itself. We have considered two kinds of trajectories.
In the first one, the barriers are adiabatically accelerated according to (\ref{traj}) until reaching the
maximum velocity with $f=f(\theta=0)=f_{\rm max}$ at $\omega t=\pi$ (Fig. \ref{figu4}), from where they maintain such a velocity.
Such trajectories are primarily intended to study the behavior of the condensate under a uniform barrier velocity. 
Note that according to (\ref{dot}) we have $I^0=N\pi(\partial Z_0/\partial\theta)f$, so a uniform barrier velocity
will keep the current bias also uniform, except for nearly overlapping barriers (see Fig. \ref{figu2}). Moreover, under
such conditions we could approximate $I^0(\theta)/I^0(\theta=0)\simeq f(\theta)/f(\theta=0)$, 
from which we may 
conclude that the ordinate of Fig. \ref{figu4} can also be regarded as the bias current $I^0/I^0_{\rm max}$.
\begin{figure}
\includegraphics{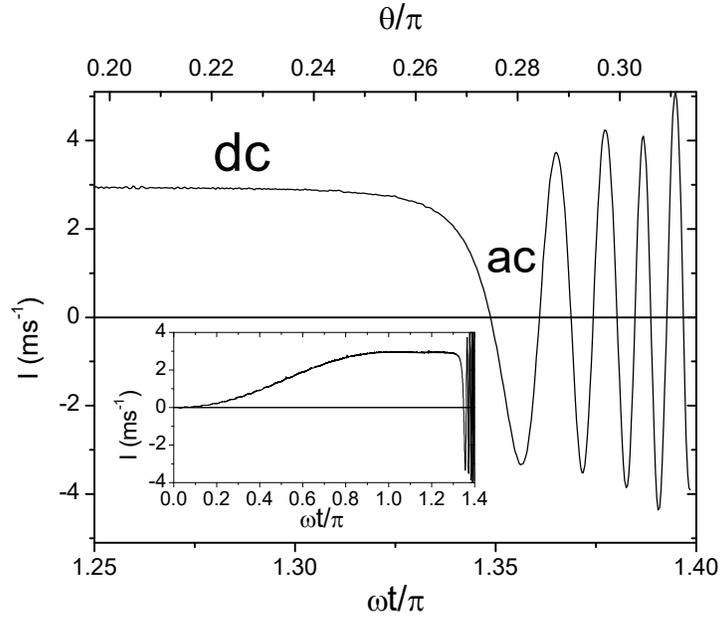}
\caption{GP simulation results for the time evolution of the
particle current for the barrier trajectory (\ref{traj}) with $f_{\rm max}=0.38$ Hz and maintaining
the maximum barrier speed for $\omega t>\pi$. The main plot shows
the dc to ac transition that occurs when $f_c(\theta(t))$ drops below $f_{\rm max}$ (cf Fig.
\ref{figu3}), with $\theta(t)$ given in the top abscissa. The inset shows the particle current for 
the complete barrier trajectory.}
\label{figu5p}
\end{figure}
We depict in Fig. \ref{figu5p} the GP simulation results for the time evolution of the
particle current for a barrier trajectory of this kind with
$f_{\rm max}=0.38$ Hz. We may observe the dc to ac transition, which occurs, despite of the fixed value of
the bias current (plateau of the inset in Fig. \ref{figu5p}),
due to the dependence of the critical current on the barrier angle, as shown in Fig. \ref{figu3}. 

\begin{figure}
\includegraphics{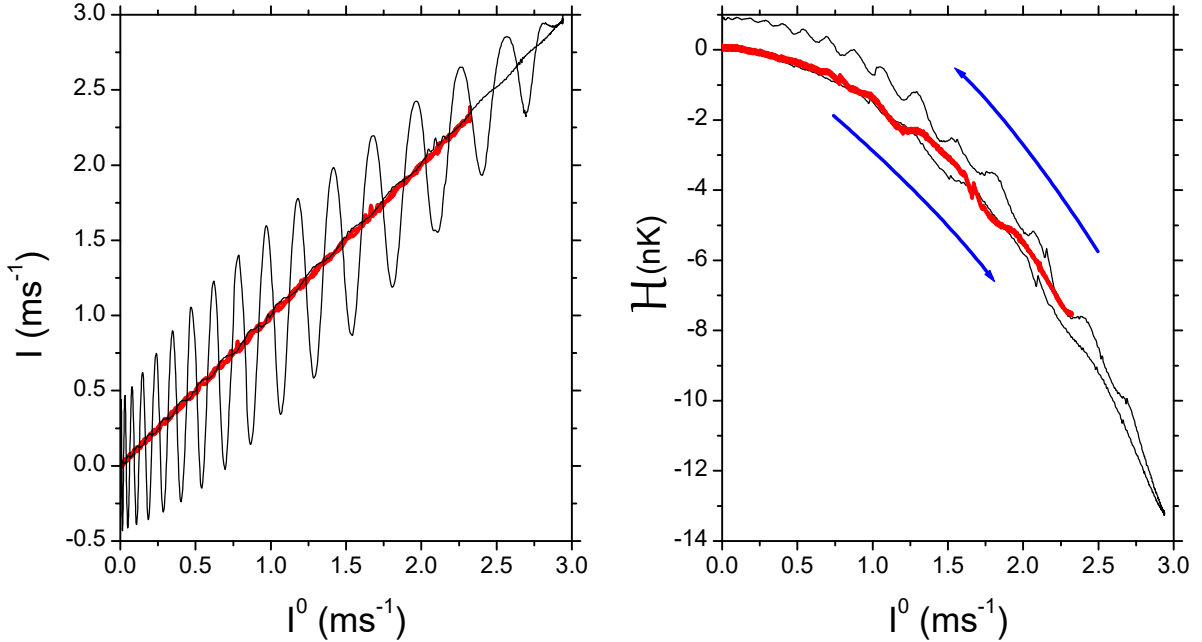}
\caption{Left panel: GP simulation results for the
particle current versus bias current ($I$ vs $I^0$) paths for the symmetric barrier trajectories of Fig. \ref{figu4}.
The non-hysteretic path corresponding to $f_{\rm max}=0.3$ Hz is depicted by the thick (red) solid line, 
while the hysteresis loop 
corresponding to $f_{\rm max}=0.3815$ Hz is represented by the thin (black) solid line. 
Right panel: same paths for the energy ${\cal H}=\Delta E_{TM}+\hbar I^0\phi$, where the blue arrows
indicate the path directions.}
\label{figu5pp}
\end{figure}

The other kind of barrier
trajectories, corresponding to the symmetric curves of Fig. \ref{figu4}, turns out to be particularly 
useful to observe hysteresis. We first note that the maximum bias
current for such trajectories (maximum barrier speed in Fig. \ref{figu4}) 
occurs for $\theta=0$, with a critical current slightly below 3 ms$^{-1}$ 
and a critical rotation frequency obtained from GP simulations of about 0.382 Hz (Fig. \ref{figu3}). 
Hysteretic and non-hysteretic evolutions are clearly represented in
the particle current versus bias current graph shown in Fig. \ref{figu5pp}.
In fact, we may see that for a barrier trajectory with $f_{\rm max}=0.3$ Hz, well below the critical value, the particle current follows
exactly the bias current for the whole path, except for the low amplitude (plasma) oscillations. On the other hand,
for $f_{\rm max}=0.3815$ Hz, although the particle current matches again the bias current
for the first half of the path,
 once the bias current reaches its maximum close to the 
critical current, it
gives rise to a quite hysteretic loop with large amplitude oscillations of the particle current. 
In the right panel of Fig. \ref{figu5pp}, we may see a similar behavior for the energy,
since for the barrier trajectory with
$f_{\rm max}=0.3$ Hz, we have fully overlapped paths for increasing and decreasing bias currents, while for
$f_{\rm max}=0.3815$ Hz,  such energy paths
become split, yielding a way back with a higher energy, which
stems from the above oscillations of the particle current. 
However, such hysteretic effects may be difficult to observe 
in practice due to the requirement of an extremely fine
tuning of the barrier velocity close to the critical value. A more favorable scenario for the hysteresis observation 
could take place for a barrier trajectory (\ref{traj}) with $f_{\rm max}>f_c$, along with
the presence of a resistive flow in
the ac regime able to bring the condensate back to the dc domain.
In addition to the effect of noncondensed atoms in a thermal component, such resistive flows may arise from
quantum phase slips corresponding to vortices created within the barrier and shed into the superfluid  \cite{jen14,xhani}. In any case, we may analyze such a scenario
by considering a suitable circuit analogy.
It is worth mentioning in this respect, that as a basic representation of the wide interrelation 
between electronics and atomtronics \cite{amico17}, 
simple models of electronic circuitry have been shown to capture 
the essential physics of superfluid transport in ultracold gases \cite{lee,eckel16,burchianti}.
Particularly, the RCSJ model \cite{stewart}, 
which consists in a very simple equivalent
circuit proven to be exceptionally successful in modeling the dynamics of superconducting Josephson devices
\cite{barone,tink,gross2016,clarke,blackburn}, has also been shown to yield
similarly good results for the dc and ac Josephson effects in ultracold gases \cite{giova00,levy,scazza}.
The RCSJ equivalent circuit for a JJ is composed of three parallel elements: a shunt resistance $R$, a 
shunt capacitance 
$C$ and a pure Josephson element that works as a nonlinear inductance \cite{martinis,gross2016}. Thus,
Kirchhoff's law corresponds in our case to $I^0 = I_s+I_n+I_d$,
where the bias current $I^0$ yields the three parallel 
currents: the superfluid
current $I_s=-I_c\sin\phi$, the normal ohmic current $I_n=-G\Delta\mu$ and the
displacement current $I_d=-C d(\Delta\mu)/dt$, being $G=1/R$ the conductance and 
\begin{eqnarray}
\Delta\mu &=& \hbar d\phi/dt \nonumber\\
&=& \Delta N_2/C,
\label{capacity}
\end{eqnarray}
where the first line (cf. Eq. (\ref{fip})) corresponds to the general voltage-phase relation of the 
Josephson effect and
the second line corresponds to the
definition of the capacitance $C$ as the ratio of the particle number difference from the equilibrium
value $\Delta N_2=N_2-N_2^0=N(Z-Z_0)/2$ 
and the chemical potential difference $\Delta\mu=\mu_2-\mu_1$. Now, replacing the displacement
current according to (\ref{capacity}) in Kirchhoff's law we obtain,
\begin{equation}
I=I_s+I_n,
\label{kirch}
\end{equation}
where $I=N\dot Z/2=\dot N_2$ denotes the particle current flowing from well `1' to well `2' (Fig. 
\ref{figu1}). Equation (\ref{kirch}) tells us that such a particle current 
consists of a superfluid component $I_s$ and a normal
component $I_n=-\Delta\mu/R$, which will only be nonnegligible for the finite chemical
potential differences of the ac regime. As regards the superfluid current $I_s=-I_c\sin\phi$, we
may obtain the expression of the Josephson inductance $L_J$ from the time derivative
$d I_s/dt=-I_c\cos\phi\, \Delta\mu/\hbar=-\Delta\mu/L_J$, which yields the phase-dependent
expression $L_J=\hbar/(I_c\cos\phi)$ \cite{martinis,gross2016}.

\begin{figure}
\includegraphics{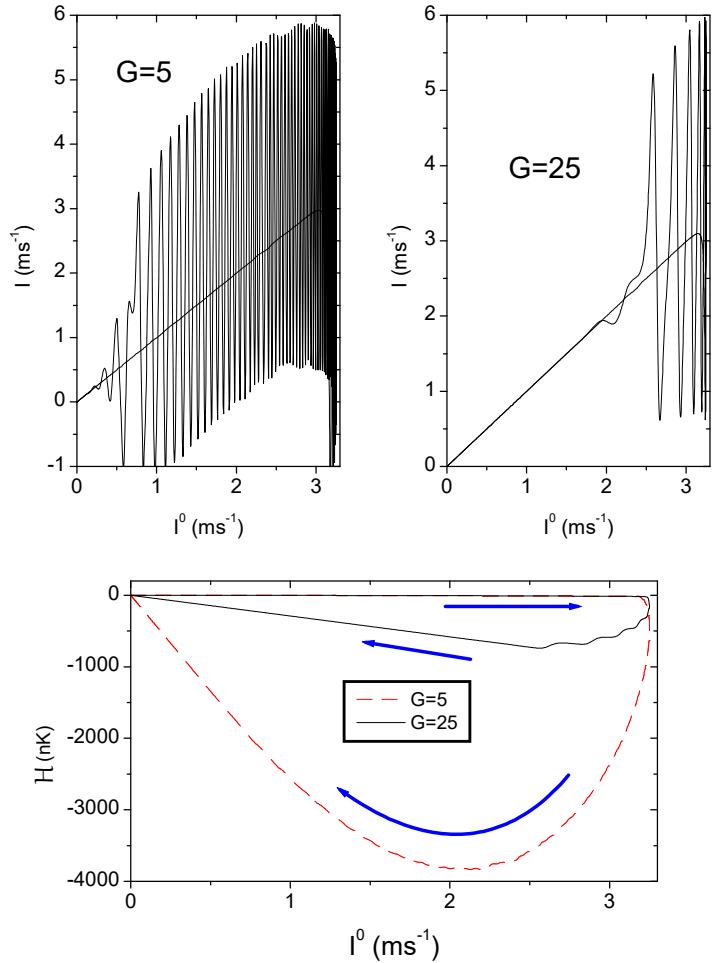}
\caption{Same as Fig. \ref{figu5pp} for the results arising from a damped TM model with
$f_{\rm max}=0.42$ Hz and two values of
the conductance $G$ (in units of $\hbar^{-1}$).}
\label{figu5ppp}
\end{figure}

Although the RCSJ model accurately
describes the experimental results in cold gases, some parameters
of the model are obtained by fitting the data and lack of a rigorous derivation. So,
a detailed comparison
with the more fundamental TM model should be relevant in this respect.
To this aim, we first compare Eq. 
(\ref{capacity}) with the TM equation (\ref{fip}), from which we may
immediately obtain the following expression for
the capacitance
\begin{equation}
C=\frac{1}{2(U+BZ)},
\end{equation}
which jointly with the Josephson inductance $L_J$ constitute an LC oscillator at the frequency
$1/\sqrt{L_J C}$, whose low amplitude limit ($\phi\simeq 0$ and $Z \simeq Z_0$) yields
the more accurate
expression $\omega_p=\sqrt{2I_c(U+BZ_0)/\hbar}$
for the plasma frequency than that given at the end of Sec. \ref{secRest}.
On the other hand, if one approximates the TM
equation (\ref{zpun}) for $Z\simeq Z_0$ and neglects the term 
proportional to the second order hopping parameter $S$, one obtains $N\dot Z/2=-I_c\sin\phi$, which
just corresponds to the Eq. (\ref{kirch}) in the dc regime. As regards
the normal component $I_n$, it is absent from
our TM model, since it stems from a formalism without any assumption about dissipative channels. 
The simplest way of including
such dissipative effects would consist in adding to the right-hand side of Eq. (\ref{zpun})
a term proportional to the RCSJ normal current $-G\Delta\mu$,
along with a phenomenological value of the conductance $G$ \cite{marino}. More elaborate procedures
that quantitatively take into account the effect of damping at a finite temperature could be carried out 
by resorting to a stochastic projected GP equation \cite{bidasyuk}.
We depict in Fig. \ref{figu5ppp} the results of the above simple version of a damped TM model 
for a symmetric barrier trajectory with
$f_{\rm max}=0.42$ Hz above the critical value and two values of the conductance.
We observe that the particle current follows the increasing values of the 
bias current up to the critical
point, from where it begins to display oscillations of a higher amplitude than those observed in Fig. \ref{figu5pp}. 
Such oscillations, which correspond to a dynamics within the ac regime, persist with the decreasing
bias current up to a value which marks the reentrance to the dc regime. This behavior 
turns out to be analogous
to the phenomenon of a return or `retrapping' current in a hysteretic superconducting JJ,
occurring when the junction switches back from the voltage state to the zero-voltage state 
\cite{martinis90,castellano}.
Although such an ac to dc transition is common in superconducting systems, 
we are not aware of any observation of this kind in a BEC.
As depicted in Fig. \ref{figu5ppp}, such a return current grows with the conductance, leading 
for high values of $G$ to an overdamped and quasi-non-hysteretic motion. In contrast, low conductances
with small return currents yield longer evolutions within the ac regime
and, hence, quite hysteretic processes. 
It is interesting to relate hysteresis with the energy evolutions depicted in Figs. 
\ref{figu5pp} and \ref{figu5ppp}. On the one hand, the system dynamics driven by the barrier trajectory with $f_{\rm max}=0.3815$ Hz in Fig. \ref{figu5pp}, which entirely takes place within the dc regime, 
develops by performing a round trip that travels twice the same line of minima of the 
energy landscape (Figs. \ref{figu3p} and \ref{figu3}), with hysteretic effects stemming
 from the oscillations around such minima developed during the return path. In contrast,
the evolutions depicted in Fig. \ref{figu5ppp} correspond to the more common hysteresis scenario
which involves more than a single minimum of the energy landscape \cite{mueller}. In fact,   
for increasing bias currents up to the critical value, the system travels the above
line of energy minima ending at the critical
point $I^0=I_c$ in Fig. \ref{figu3p}. Then, the system leaves the dc regime 
and entering
the ac domain, there is an onset of 
a `running down hill' process due to the absence of any local energy minimum. Next,
following the subsequent 
barrier trajectory, the decreasing bias current falls again below the critical value with
a reappearance of local energy minima and the possibility for the system to be `retrapped'
and thus able to return to the dc domain and travel a different line of minima than that 
left after entering the ac regime. We depict in the lower panel of Fig. \ref{figu5ppp} the hysteresis
loops performed by the energy versus bias current trajectories for two values of the conductance $G$.
Note that the smaller the conductance, the deeper energy fall 
occurring as the system remains in the ac regime.
However, for conductance values below certain threshold, the return to the dc domain does not occur,
and in the final configuration of
a vanishing bias current (barriers at rest), the system ends in a high-energy 
nonequilibrium self-trapped state, with a quite compressed condensate at site 1 (Fig. \ref{figu1}) \cite{boshier13,jen14,cat14}. Finally, we remark that it would be interesting to generate experimental 
results about the return transition from ac to dc, which would 
allow to test the validity of the above predictions of the damped TM model.

\subsubsection{GP simulations and comparison to TM model results}\label{secGPTM}
\begin{figure}
\includegraphics{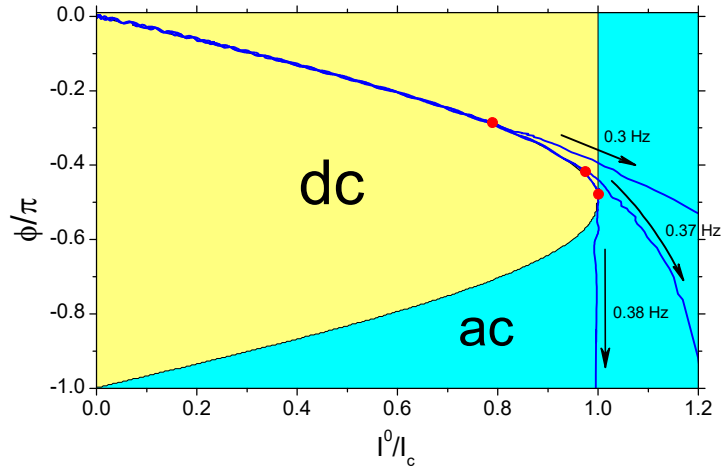}
\caption{Phase difference versus bias current from GP simulation results
 for three barrier trajectories
(\ref{traj}) with constant values of 0.3, 0.37 and 0.38 Hz of the barrier rotation frequency from the corresponding
red dots.}
\label{figu5}
\end{figure}
\begin{figure}
\includegraphics{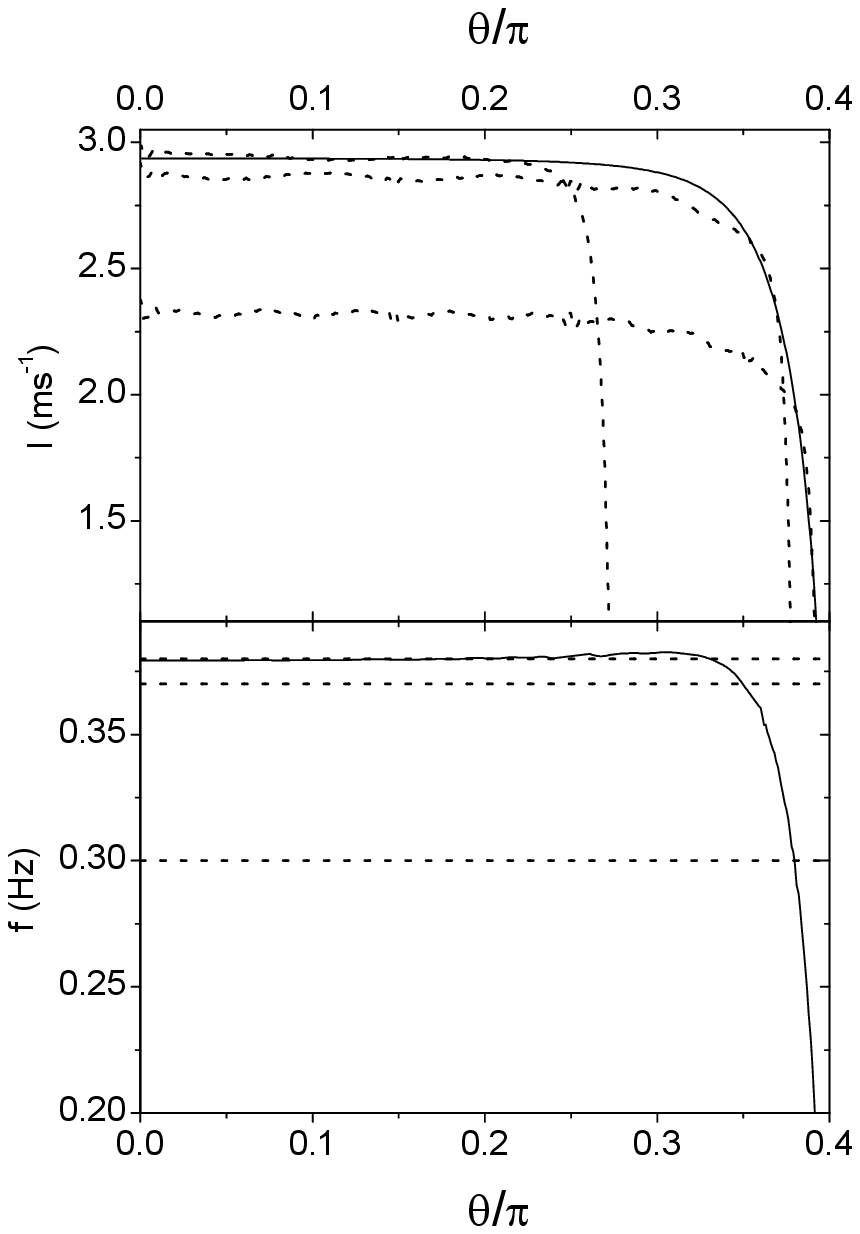}
\caption{Top panel:  critical current $I_c$  versus the barrier angle $\theta$
(solid line) and particle currents arising from GP simulations (dotted lines) 
for the barrier trajectories of Fig. \ref{figu5} with the uniform rotation frequencies
0.38, 0.37 and 0.3 Hz attained from $\theta=0$ with, respectively,
top to bottom intersections with the ordinate and left to right intersections with the abscissa.
The flow oscillations of the ac regime have not been displayed for clarity.
Bottom panel: same as top panel for the critical rotation frequency $f_c$ (solid line) as compared
to the rotation frequencies 0.38, 0.37 and 0.3 Hz (dotted lines) from top to bottom, respectively.}
\label{figu9p}
\end{figure}
We have studied the transition from the dc to the ac regime
for different barrier trajectories and initial states of the system. Unless stated, we will assume an initial condensate
in equilibrium at the ground state.
First we will consider the evolution for a uniform barrier velocity after the initial adiabatic
acceleration. 
We depict in Fig. \ref{figu5} the phase difference versus bias current for three values of the final barrier velocity.
We may observe that the system travels quite closely the line of energy minima 
(Fig. \ref{figu3}, inset) up to reaching 
the uniform rotation frequency (red dots). Then, the subsequent evolution in the dc regime is 
represented by the red dots in Fig. \ref{figu5}, since the condensate stays with a constant
phase difference and a uniform particle/bias current. Finally, a sudden transition to
the ac regime occurs when the Josephson critical current becomes smaller than the particle/bias current
for the increasing barrier angles (Fig. \ref{figu3}). In fact, the top panel of Fig. \ref{figu9p}
displays such a behavior for the particle currents corresponding to the barrier trajectories with 
$f_{\rm max}=0.3$ Hz and $f_{\rm max}=0.37$ Hz, while the bottom panel also shows that the
critical condition for the barrier rotation frequency $f = f_c$ happens quite simultaneously with the
critical crossings $I = I_c$ in the top panel.
On the other hand, the case $f_{\rm max}=0.38$ Hz shows important differences with respect to
those of the lower rotation frequencies. We notice in Fig. \ref{figu9p} that the steady particle current
for $\theta >0$ obtained from the GP simulation, as well as the rotation frequency 
0.38 Hz, practically coincide with the critical values arising from the TM model for $\theta/\pi\lesssim
0.2$.
Thus, it is easy to understand that under such conditions the model should not be expected to yield accurate
results for the barrier angle at which the transition should occur. 
\begin{figure}
\includegraphics{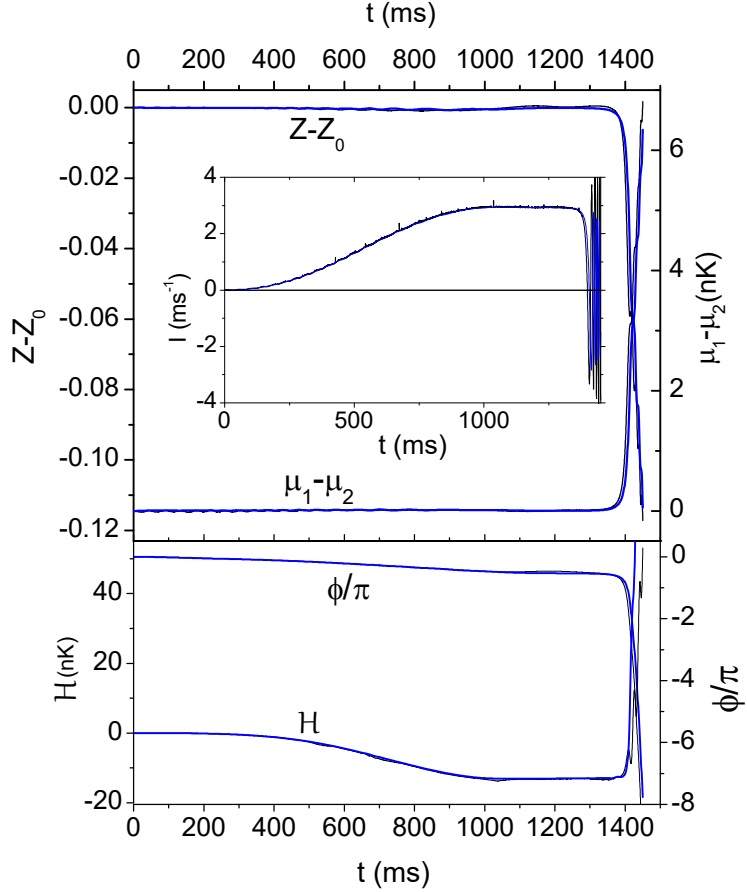}
\caption{Time evolution of the imbalance departure from the equilibrium value $Z-Z_0$, chemical potencial
difference $\mu_1-\mu_2$, particle current $I$ (inset),
phase difference $\phi$,  and energy ${\cal H}$ from GP simulation results (black solid lines)
for the barrier motion with a maximum rotation frequency $f_{\rm max}=0.38$ Hz referred to in 
Figs. \ref{figu5p}, \ref{figu5} and \ref{figu9p}. The blue solid lines represent the corresponding
TM model results with a best fit value of $f_{\rm max}=0.37862$ Hz.}
\label{figu5e}
\end{figure}

In Fig. \ref{figu5e}, we may appreciate the way in which
such a transition is reflected on the condensate evolution with the highest 
rotation frequency $f_{\rm max}=0.38$ Hz of Fig. \ref{figu5}. In addition to the plain dc-ac 
current transition already
shown in Fig. \ref{figu5p}, we may observe
that the remaining condensate observables show quite sharp variations at the transition. 
In fact, the sudden compression undergone by the condensate at site 1, along 
with the corresponding expansion at site 2, becomes reflected in the top panel of Fig. \ref{figu5e} through
the sharp decrease (increase) of the imbalance departure from the equilibrium value 
$Z-Z_0$ (chemical potential difference $\mu_1-\mu_2$) at the transition 
from the almost vanishing values shown in the dc regime. On the other hand, 
the bottom panel of Fig. \ref{figu5e} shows that after the adiabatic barrier acceleration
(below 1000 ms), the phase difference and the energy remain constant up to the transition to the ac regime, 
which is characterized by a running downwards phase difference and an energy jump stemming
from the sudden compression/expansion of the condensate 
that triggers the term $\sim (Z-Z_0)^2$ in (\ref{ener}). Here it is important to remark
the excellent agreement 
that show in Fig. \ref{figu5e} the GP simulation results and the corresponding TM results 
for a slightly modified best fit value of the maximum rotation frequency (less than 0.4\%),
with respect to that of the GP simulation.

\begin{figure}
\includegraphics{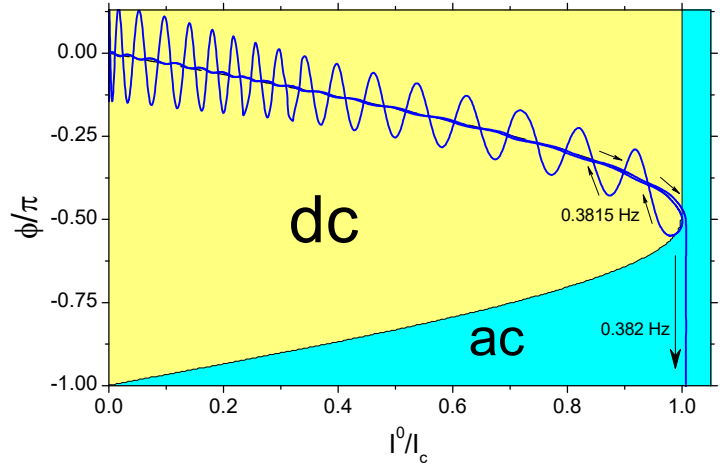}
\caption{GP simulation results for the phase difference versus bias current for two symmetric barrier trajectories
(Fig. \ref{figu4}) with $f_{\rm max}=0.3815$ Hz and $f_{\rm max}=0.382$ Hz.}
\label{figu6}
\end{figure}

\begin{figure}
\includegraphics{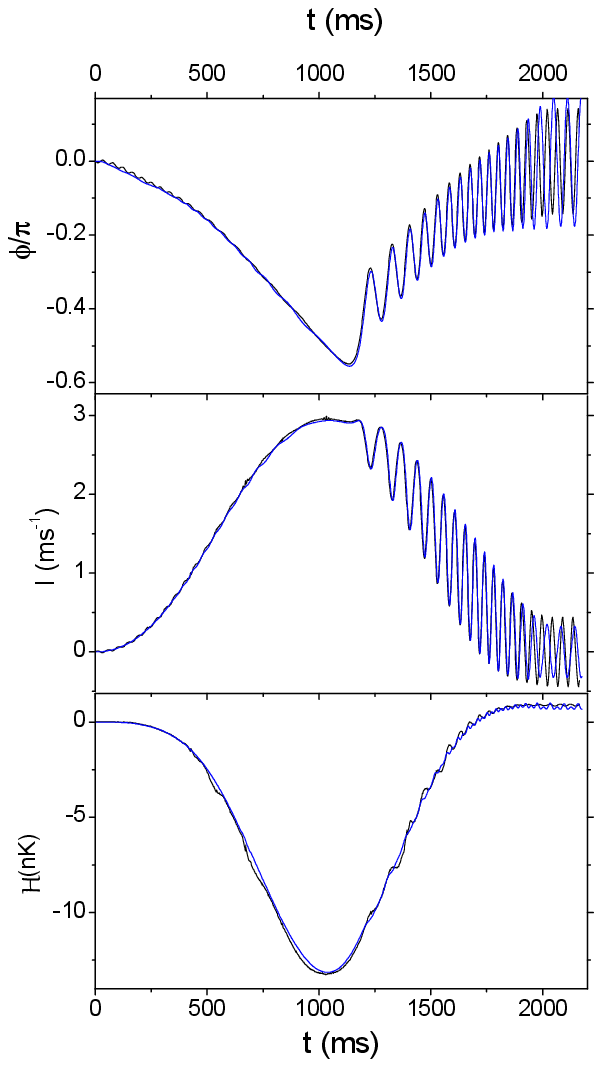}
\caption{Time evolution of the phase difference $\phi$, particle current $I$ and energy ${\cal H}$ from GP simulation
results (black solid lines) and TM model results (blue solid lines) 
for symmetric barrier trajectories (Fig. \ref{figu4}) almost touching the critical point,
with $f_{\rm max}=0.3815$ Hz (GP simulation) and 
$f_{\rm max}=0.37958$ Hz (TM model).}
\label{figu6e}
\end{figure}

The condensate evolution for two slightly different symmetric trajectories of the barriers is depicted in Fig. \ref{figu6}.
Here the faster trajectory with $f_{\rm max}=0.382$ Hz provokes the condensate transition to the ac regime, while the
slower trajectory ($f_{\rm max}=0.3815$ Hz) yields the hysteretic loop within the dc domain, 
already seen in  Fig. \ref{figu5pp}. It is interesting to observe in Fig. \ref{figu6} that both trajectories in the phase
difference versus bias current plane share the first part of the path, as they travel the line of energy minima up to the
critical point, where minimum and saddle coalesce. At this point the condensate driven by the faster barriers `drops'
to the ac domain, while that driven by the slower barriers go a little further along the line of energy saddles. 
Note in Fig. \ref{figu6} that this is precisely the
fact that makes the return path (decreasing bias currents) oscillate around the line of energy minima. In other
words, such a `tour' beyond the critical point should be regarded as the source of the hysteretic behavior.
To pursue with the study of this case, we depict in Fig. \ref{figu6e} time evolutions of the phase difference,
particle current and energy that complement what represented in Figs. \ref{figu5pp} and \ref{figu6}. Again, an
excellent agreement between GP simulation and TM results is obtained with a slightly less best fit
value of the maximum rotation frequency for the TM model (0.5\%).

\begin{figure}
\includegraphics{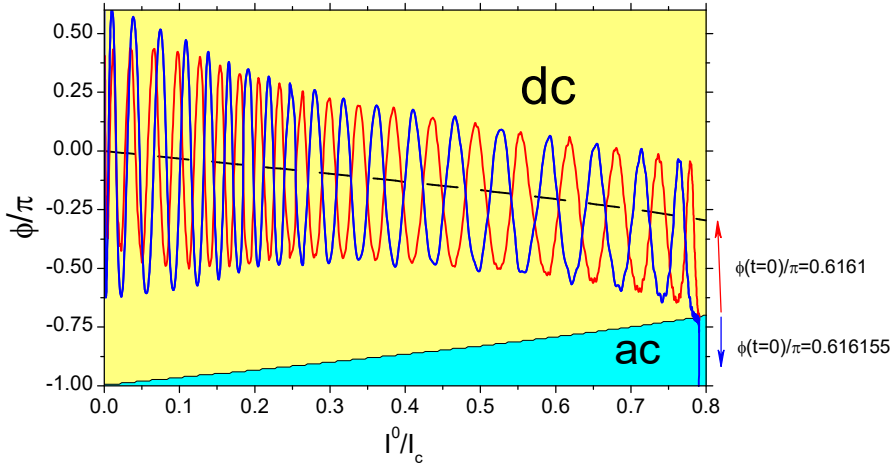}
\caption{Phase difference versus bias current from GP simulation results
for two symmetric barrier trajectories
(Fig. \ref{figu4}) with $f_{\rm max}=0.3$ Hz and nonvanishing initial phase differences of $0.6161\pi$
and $0.616155\pi$. Both curves turn out to be undistinguishable 
within the plot scale up to the maximum bias current, 
from where the return path for the former is depicted with the red solid line, while the latter conserves the
original trace. The dashed line 
corresponds to the phase difference $\phi_m$ of the energy minimum.}
\label{figu7}
\end{figure}

\begin{figure}
\includegraphics{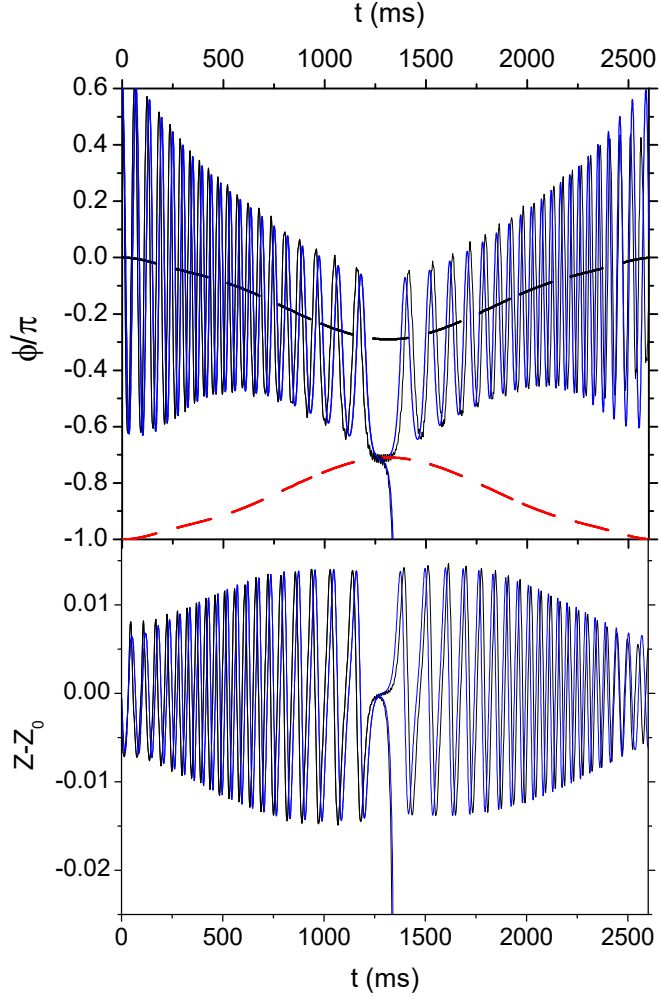}
\caption{Same as Fig. \ref{figu7} for the time evolution of the phase difference $\phi$ and the imbalance departure 
from the equilibrium value $Z-Z_0$ (black solid lines). The blue solid lines depict the corresponding
TM model results for initial phase differences of $0.6245\pi$ and $0.6248\pi$, while
the black (red) dashed line corresponds to 
the phase difference $\phi_m$ ($\phi_M$) of the energy minimum (saddle).}
\label{figu14}
\end{figure}

Finally, we will consider a different initial condition from that previously assumed of 
the ground state of the condensate.
In fact, we
will assume an initial order parameter of the form $e^{i\phi}\sqrt{(1-Z_0)/2}\,\psi_1
+\sqrt{(1+Z_0)/2}\,\psi_2$, 
with $\psi_j$ denoting the wavefunction of the localized state
on the $j$-well and the parameter $\phi$ serving to introduce an initial phase difference between both wells. 
Note that $\phi=0$ corresponds to the ground-state order parameter. 
Then, assuming again symmetric barrier trajectories, with $f_{\rm max}<f_c$ in this case, 
there will be a nonvanishing critical value of the initial phase difference above which the system will make the transition to the ac regime.
In fact, we depict in Fig. \ref{figu7} the GP simulation results for two initial phase differences just above and below such a critical
value for a barrier trajectory with $f_{\rm max}=0.3$ Hz. Thus, we may see overlapping 
phase differences for both initial conditions at increasing bias currents, yielding
a trajectory of oscillations around the line of energy minima up to reaching the maximum rotation frequency
(bias current). Next, the trajectories split into a returning oscillating path, similar to that developed for increasing bias currents
(except for a phase displacement of $\pi$, approximately), and,
for the higher initial phase difference, a trajectory that makes the transition to ac across the line of energy saddles.
In Fig. \ref{figu14}, we depict the time evolution of the phase difference and the imbalance departure from the 
equilibrium value for the above configurations, where a very good agreement is observed between the GP simulation
results and the TM model results with slightly higher best fit values of the initial phase differences ($\sim$ 1\%). 

\section{Conclusion}\label{secConclusion}
We have analized the effects of the barrier motion on the BH Hamiltonian and the equations of
 motion of an atom dc-SQUID. We have found that a couple of terms
arising from the additional particle flow induced by the barriers displacement should be added
to the condensate energy. In fact,
in addition to the well-known contribution proportional to the bias current and the phase difference, which
yields the tilting of the washboard potential, we have identified a hysteretic term that should be 
considered for accelerated or overlapping barriers. Based on such energies, we have proposed two
corresponding additional contributions to the BH Hamiltonian and have analyzed their effects on the
Heisenberg equation of motion for the boson field operator in the TM approximation. Thus,
we have found that the nonhysteretic additional contribution to the Heisenberg equation does cancel with
that arising from the time derivative of the boson field operator in the Schr\"odinger representation, whereas
the hysteretic
contribution seems to be negligible in the present case of adiabatic barrier accelerations, 
according to the experimental results. So, we have utilized formally the same GP and TM model
equations derived for barriers at rest,
except for the time dependence of parameters due to the barrier motion.
By deriving the expression of the condensate
energy from the BH Hamiltonian, we have studied the energy landscape as a function of
an order parameter (phase difference) and a control parameter (bias current), determining the
diagram with the location of dc and ac regimes. In addition, we have found that the Josephson critical current, and correspondingly the critical barrier speed, 
depend on the barrier position, a fact that makes that
a dc to ac transition could always be reached for
any uniform barrier velocity attained after the initial adiabatic acceleration. 
On the other hand, we have studied the condensate
evolution driven by barrier trajectories symmetric with respect to the 
symmetric configuration of the dc-SQUID. Particularly, we analyzed the hysteretic effects
stemming from a trajectory that almost reaches the critical point and develops an oscillating return 
within the dc domain, as
compared to a sligthly faster trajectory that yields the condensate 
transition to the ac regime. In addition, we discussed an
easier to detect hysteresis scenario that could arise for a sufficiently resistive flow in the ac regime.
We have also seen that when the initial condensate presents a phase difference between both wells, instead
of the ground state, there exists a critical value of such a phase difference above which
the dc-SQUID makes a transition to the ac regime, irrespective of the maximum value of the barrier velocity 
attained at the symmetric trajectory. To conclude, we may remark that the excellent agreement
between the GP simulation results and the TM model results found in all cases,
leaves a good open window to further apply such a model
and its corresponding BH Hamiltonian in order to gain better insights about this kind of critical phenomena.

\acknowledgments
This work was supported by grants PIP 11220150100442CO from CONICET and
UBA-CyT 20020150100157 from Universidad de Buenos Aires. Helpful discussions with
D. M. Jezek are gratefully acknowledged.

%\bibliography{paper}

\begin{thebibliography}{43}%
\makeatletter
\providecommand \@ifxundefined [1]{%
 \@ifx{#1\undefined}
}%
\providecommand \@ifnum [1]{%
 \ifnum #1\expandafter \@firstoftwo
 \else \expandafter \@secondoftwo
 \fi
}%
\providecommand \@ifx [1]{%
 \ifx #1\expandafter \@firstoftwo
 \else \expandafter \@secondoftwo
 \fi
}%
\providecommand \natexlab [1]{#1}%
\providecommand \enquote  [1]{``#1''}%
\providecommand \bibnamefont  [1]{#1}%
\providecommand \bibfnamefont [1]{#1}%
\providecommand \citenamefont [1]{#1}%
\providecommand \href@noop [0]{\@secondoftwo}%
\providecommand \href [0]{\begingroup \@sanitize@url \@href}%
\providecommand \@href[1]{\@@startlink{#1}\@@href}%
\providecommand \@@href[1]{\endgroup#1\@@endlink}%
\providecommand \@sanitize@url [0]{\catcode `\\12\catcode `\$12\catcode
  `\&12\catcode `\#12\catcode `\^12\catcode `\_12\catcode `\%12\relax}%
\providecommand \@@startlink[1]{}%
\providecommand \@@endlink[0]{}%
\providecommand \url  [0]{\begingroup\@sanitize@url \@url }%
\providecommand \@url [1]{\endgroup\@href {#1}{\urlprefix }}%
\providecommand \urlprefix  [0]{URL }%
\providecommand \Eprint [0]{\href }%
\providecommand \doibase [0]{http://dx.doi.org/}%
\providecommand \selectlanguage [0]{\@gobble}%
\providecommand \bibinfo  [0]{\@secondoftwo}%
\providecommand \bibfield  [0]{\@secondoftwo}%
\providecommand \translation [1]{[#1]}%
\providecommand \BibitemOpen [0]{}%
\providecommand \bibitemStop [0]{}%
\providecommand \bibitemNoStop [0]{.\EOS\space}%
\providecommand \EOS [0]{\spacefactor3000\relax}%
\providecommand \BibitemShut  [1]{\csname bibitem#1\endcsname}%
\let\auto@bib@innerbib\@empty
%</preamble>
\bibitem [{\citenamefont {Clarke}\ and\ \citenamefont
  {Braginski}(2004)}]{braginski}%
  \BibitemOpen
  \bibfield  {author} {\bibinfo {author} {\bibfnamefont {J.}~\bibnamefont
  {Clarke}}\ and\ \bibinfo {author} {\bibfnamefont {A.~I.}\ \bibnamefont
  {Braginski}},\ }\href@noop {} {\emph {\bibinfo {title} {The SQUID
  Handbook}}}\ (\bibinfo  {publisher} {Wiley-VCH},\ \bibinfo {address}
  {Weinheim},\ \bibinfo {year} {2004})\BibitemShut {NoStop}%
\bibitem [{\citenamefont {Sato}\ and\ \citenamefont {Packard}(2012)}]{satop}%
  \BibitemOpen
  \bibfield  {author} {\bibinfo {author} {\bibfnamefont {Y.}~\bibnamefont
  {Sato}}\ and\ \bibinfo {author} {\bibfnamefont {R.~E.}\ \bibnamefont
  {Packard}},\ }\href@noop {} {\bibfield  {journal} {\bibinfo  {journal} {Rep.
  Prog. Phys.}\ }\textbf {\bibinfo {volume} {75}},\ \bibinfo {pages} {016401}
  (\bibinfo {year} {2012})}\BibitemShut {NoStop}%
\bibitem [{\citenamefont {Ryu}\ \emph {et~al.}(2013)\citenamefont {Ryu},
  \citenamefont {Blackburn}, \citenamefont {Blinova},\ and\ \citenamefont
  {Boshier}}]{boshier13}%
  \BibitemOpen
  \bibfield  {author} {\bibinfo {author} {\bibfnamefont {C.}~\bibnamefont
  {Ryu}}, \bibinfo {author} {\bibfnamefont {P.~W.}\ \bibnamefont {Blackburn}},
  \bibinfo {author} {\bibfnamefont {A.~A.}\ \bibnamefont {Blinova}}, \ and\
  \bibinfo {author} {\bibfnamefont {M.~G.}\ \bibnamefont {Boshier}},\
  }\href@noop {} {\bibfield  {journal} {\bibinfo  {journal} {Phys. Rev. Lett.}\
  }\textbf {\bibinfo {volume} {111}},\ \bibinfo {pages} {205301} (\bibinfo
  {year} {2013})}\BibitemShut {NoStop}%
\bibitem [{\citenamefont {Sato}(2013)}]{sato}%
  \BibitemOpen
  \bibfield  {author} {\bibinfo {author} {\bibfnamefont {Y.}~\bibnamefont
  {Sato}},\ }\href@noop {} {\bibfield  {journal} {\bibinfo  {journal}
  {Physics}\ }\textbf {\bibinfo {volume} {6}},\ \bibinfo {pages} {123}
  (\bibinfo {year} {2013})}\BibitemShut {NoStop}%
\bibitem [{\citenamefont {Sackett}(2014)}]{sackett}%
  \BibitemOpen
  \bibfield  {author} {\bibinfo {author} {\bibfnamefont {C.~A.}\ \bibnamefont
  {Sackett}},\ }\href@noop {} {\bibfield  {journal} {\bibinfo  {journal}
  {Nature}\ }\textbf {\bibinfo {volume} {505}},\ \bibinfo {pages} {166}
  (\bibinfo {year} {2014})}\BibitemShut {NoStop}%
\bibitem [{\citenamefont {Barone}\ and\ \citenamefont
  {Patern\`o}(1982)}]{barone}%
  \BibitemOpen
  \bibfield  {author} {\bibinfo {author} {\bibfnamefont {A.}~\bibnamefont
  {Barone}}\ and\ \bibinfo {author} {\bibfnamefont {G.}~\bibnamefont
  {Patern\`o}},\ }\href@noop {} {\emph {\bibinfo {title} {Physics and
  Applications of the Josephson Effect}}}\ (\bibinfo  {publisher} {John
  Wiley},\ \bibinfo {address} {New York},\ \bibinfo {year} {1982})\BibitemShut
  {NoStop}%
\bibitem [{\citenamefont {Tinkham}(1996)}]{tink}%
  \BibitemOpen
  \bibfield  {author} {\bibinfo {author} {\bibfnamefont {M.}~\bibnamefont
  {Tinkham}},\ }\href@noop {} {\emph {\bibinfo {title} {Introduction to
  Superconductivity}}},\ \bibinfo {edition} {2nd}\ ed.\ (\bibinfo  {publisher}
  {McGraw-Hill},\ \bibinfo {address} {New York},\ \bibinfo {year}
  {1996})\BibitemShut {NoStop}%
\bibitem [{\citenamefont {Gross}\ \emph {et~al.}(2016)\citenamefont {Gross},
  \citenamefont {Marx},\ and\ \citenamefont {Deppe}}]{gross2016}%
  \BibitemOpen
  \bibfield  {author} {\bibinfo {author} {\bibfnamefont {R.}~\bibnamefont
  {Gross}}, \bibinfo {author} {\bibfnamefont {A.}~\bibnamefont {Marx}}, \ and\
  \bibinfo {author} {\bibfnamefont {F.}~\bibnamefont {Deppe}},\ }\href@noop {}
  {\emph {\bibinfo {title} {Applied Superconductivity: Josephson Effect and
  Superconducting Electronics}}},\ De Gruyter Textbook Series\ (\bibinfo
  {publisher} {Walter De Gruyter},\ \bibinfo {address} {Berlin},\ \bibinfo
  {year} {2016})\BibitemShut {NoStop}%
\bibitem [{\citenamefont {Giovanazzi}\ \emph {et~al.}(2000)\citenamefont
  {Giovanazzi}, \citenamefont {Smerzi},\ and\ \citenamefont
  {Fantoni}}]{giova00}%
  \BibitemOpen
  \bibfield  {author} {\bibinfo {author} {\bibfnamefont {S.}~\bibnamefont
  {Giovanazzi}}, \bibinfo {author} {\bibfnamefont {A.}~\bibnamefont {Smerzi}},
  \ and\ \bibinfo {author} {\bibfnamefont {S.}~\bibnamefont {Fantoni}},\
  }\href@noop {} {\bibfield  {journal} {\bibinfo  {journal} {Phys. Rev. Lett.}\
  }\textbf {\bibinfo {volume} {84}},\ \bibinfo {pages} {4521} (\bibinfo {year}
  {2000})}\BibitemShut {NoStop}%
\bibitem [{\citenamefont {Levy}\ \emph {et~al.}(2007)\citenamefont {Levy},
  \citenamefont {Lahoud}, \citenamefont {Shomroni},\ and\ \citenamefont
  {Steinhauer}}]{levy}%
  \BibitemOpen
  \bibfield  {author} {\bibinfo {author} {\bibfnamefont {S.}~\bibnamefont
  {Levy}}, \bibinfo {author} {\bibfnamefont {E.}~\bibnamefont {Lahoud}},
  \bibinfo {author} {\bibfnamefont {I.}~\bibnamefont {Shomroni}}, \ and\
  \bibinfo {author} {\bibfnamefont {J.}~\bibnamefont {Steinhauer}},\
  }\href@noop {} {\bibfield  {journal} {\bibinfo  {journal} {Nature}\ }\textbf
  {\bibinfo {volume} {449}},\ \bibinfo {pages} {579} (\bibinfo {year}
  {2007})}\BibitemShut {NoStop}%
\bibitem [{\citenamefont {Jendrzejewski}\ \emph {et~al.}(2014)\citenamefont
  {Jendrzejewski}, \citenamefont {Eckel}, \citenamefont {Murray}, \citenamefont
  {Lanier}, \citenamefont {Edwards}, \citenamefont {Lobb},\ and\ \citenamefont
  {Campbell}}]{jen14}%
  \BibitemOpen
  \bibfield  {author} {\bibinfo {author} {\bibfnamefont {F.}~\bibnamefont
  {Jendrzejewski}}, \bibinfo {author} {\bibfnamefont {S.}~\bibnamefont
  {Eckel}}, \bibinfo {author} {\bibfnamefont {N.}~\bibnamefont {Murray}},
  \bibinfo {author} {\bibfnamefont {C.}~\bibnamefont {Lanier}}, \bibinfo
  {author} {\bibfnamefont {M.}~\bibnamefont {Edwards}}, \bibinfo {author}
  {\bibfnamefont {C.~J.}\ \bibnamefont {Lobb}}, \ and\ \bibinfo {author}
  {\bibfnamefont {G.~K.}\ \bibnamefont {Campbell}},\ }\href@noop {} {\bibfield
  {journal} {\bibinfo  {journal} {Phys. Rev. Lett.}\ }\textbf {\bibinfo
  {volume} {113}},\ \bibinfo {pages} {045305} (\bibinfo {year}
  {2014})}\BibitemShut {NoStop}%
\bibitem [{\citenamefont {Wright}\ \emph {et~al.}(2013)\citenamefont {Wright},
  \citenamefont {Blakestad}, \citenamefont {Lobb}, \citenamefont {Phillips},\
  and\ \citenamefont {Campbell}}]{wright}%
  \BibitemOpen
  \bibfield  {author} {\bibinfo {author} {\bibfnamefont {K.~C.}\ \bibnamefont
  {Wright}}, \bibinfo {author} {\bibfnamefont {R.~B.}\ \bibnamefont
  {Blakestad}}, \bibinfo {author} {\bibfnamefont {C.~J.}\ \bibnamefont {Lobb}},
  \bibinfo {author} {\bibfnamefont {W.~D.}\ \bibnamefont {Phillips}}, \ and\
  \bibinfo {author} {\bibfnamefont {G.~K.}\ \bibnamefont {Campbell}},\
  }\href@noop {} {\bibfield  {journal} {\bibinfo  {journal} {Phys. Rev. Lett.}\
  }\textbf {\bibinfo {volume} {110}},\ \bibinfo {pages} {025302} (\bibinfo
  {year} {2013})}\BibitemShut {NoStop}%
\bibitem [{\citenamefont {Eckel}\ \emph {et~al.}(2014)\citenamefont {Eckel},
  \citenamefont {Lee}, \citenamefont {Jendrzejewski}, \citenamefont {Murray},
  \citenamefont {Clark}, \citenamefont {Lobb}, \citenamefont {Phillips},
  \citenamefont {Edwards},\ and\ \citenamefont {Campbell}}]{eckel}%
  \BibitemOpen
  \bibfield  {author} {\bibinfo {author} {\bibfnamefont {S.}~\bibnamefont
  {Eckel}}, \bibinfo {author} {\bibfnamefont {J.~G.}\ \bibnamefont {Lee}},
  \bibinfo {author} {\bibfnamefont {F.}~\bibnamefont {Jendrzejewski}}, \bibinfo
  {author} {\bibfnamefont {N.}~\bibnamefont {Murray}}, \bibinfo {author}
  {\bibfnamefont {C.~W.}\ \bibnamefont {Clark}}, \bibinfo {author}
  {\bibfnamefont {C.~J.}\ \bibnamefont {Lobb}}, \bibinfo {author}
  {\bibfnamefont {W.~D.}\ \bibnamefont {Phillips}}, \bibinfo {author}
  {\bibfnamefont {M.}~\bibnamefont {Edwards}}, \ and\ \bibinfo {author}
  {\bibfnamefont {G.~K.}\ \bibnamefont {Campbell}},\ }\href {\doibase
  10.1038/nature12958} {\bibfield  {journal} {\bibinfo  {journal} {Nature}\
  }\textbf {\bibinfo {volume} {506}},\ \bibinfo {pages} {200} (\bibinfo {year}
  {2014})}\BibitemShut {NoStop}%
\bibitem [{\citenamefont {Cataldo}\ and\ \citenamefont {Jezek}(2014)}]{cat14}%
  \BibitemOpen
  \bibfield  {author} {\bibinfo {author} {\bibfnamefont {H.~M.}\ \bibnamefont
  {Cataldo}}\ and\ \bibinfo {author} {\bibfnamefont {D.~M.}\ \bibnamefont
  {Jezek}},\ }\href@noop {} {\bibfield  {journal} {\bibinfo  {journal} {Phys.
  Rev. A}\ }\textbf {\bibinfo {volume} {90}},\ \bibinfo {pages} {043610}
  (\bibinfo {year} {2014})}\BibitemShut {NoStop}%
\bibitem [{\citenamefont {Jezek}\ \emph {et~al.}(2013)\citenamefont {Jezek},
  \citenamefont {Capuzzi},\ and\ \citenamefont {Cataldo}}]{cap13}%
  \BibitemOpen
  \bibfield  {author} {\bibinfo {author} {\bibfnamefont {D.~M.}\ \bibnamefont
  {Jezek}}, \bibinfo {author} {\bibfnamefont {P.}~\bibnamefont {Capuzzi}}, \
  and\ \bibinfo {author} {\bibfnamefont {H.~M.}\ \bibnamefont {Cataldo}},\
  }\href@noop {} {\bibfield  {journal} {\bibinfo  {journal} {Phys. Rev. A}\
  }\textbf {\bibinfo {volume} {87}},\ \bibinfo {pages} {053625} (\bibinfo
  {year} {2013})}\BibitemShut {NoStop}%
\bibitem [{\citenamefont {Jezek}\ and\ \citenamefont {Cataldo}(2013)}]{je13}%
  \BibitemOpen
  \bibfield  {author} {\bibinfo {author} {\bibfnamefont {D.~M.}\ \bibnamefont
  {Jezek}}\ and\ \bibinfo {author} {\bibfnamefont {H.~M.}\ \bibnamefont
  {Cataldo}},\ }\href@noop {} {\bibfield  {journal} {\bibinfo  {journal} {Phys.
  Rev. A}\ }\textbf {\bibinfo {volume} {88}},\ \bibinfo {pages} {013636}
  (\bibinfo {year} {2013})}\BibitemShut {NoStop}%
\bibitem [{\citenamefont {Kautz}\ and\ \citenamefont
  {Martinis}(1990)}]{martinis90}%
  \BibitemOpen
  \bibfield  {author} {\bibinfo {author} {\bibfnamefont {R.~L.}\ \bibnamefont
  {Kautz}}\ and\ \bibinfo {author} {\bibfnamefont {J.~M.}\ \bibnamefont
  {Martinis}},\ }\href@noop {} {\bibfield  {journal} {\bibinfo  {journal}
  {Phys. Rev. B}\ }\textbf {\bibinfo {volume} {42}},\ \bibinfo {pages} {9903}
  (\bibinfo {year} {1990})}\BibitemShut {NoStop}%
\bibitem [{\citenamefont {Castellano}\ \emph {et~al.}(1999)\citenamefont
  {Castellano}, \citenamefont {Torrioli}, \citenamefont {Chiarello},
  \citenamefont {Cosmelli},\ and\ \citenamefont {Carelli}}]{castellano}%
  \BibitemOpen
  \bibfield  {author} {\bibinfo {author} {\bibfnamefont {M.~G.}\ \bibnamefont
  {Castellano}}, \bibinfo {author} {\bibfnamefont {G.}~\bibnamefont
  {Torrioli}}, \bibinfo {author} {\bibfnamefont {F.}~\bibnamefont {Chiarello}},
  \bibinfo {author} {\bibfnamefont {C.}~\bibnamefont {Cosmelli}}, \ and\
  \bibinfo {author} {\bibfnamefont {P.}~\bibnamefont {Carelli}},\ }\href@noop
  {} {\bibfield  {journal} {\bibinfo  {journal} {J. Appl. Phys.}\ }\textbf
  {\bibinfo {volume} {86}},\ \bibinfo {pages} {6405} (\bibinfo {year}
  {1999})}\BibitemShut {NoStop}%
\bibitem [{\citenamefont {Stewart}(1968)}]{stewart}%
  \BibitemOpen
  \bibfield  {author} {\bibinfo {author} {\bibfnamefont {W.~C.}\ \bibnamefont
  {Stewart}},\ }\href@noop {} {\bibfield  {journal} {\bibinfo  {journal} {Appl.
  Phys. Lett.}\ }\textbf {\bibinfo {volume} {12}},\ \bibinfo {pages} {277}
  (\bibinfo {year} {1968})}\BibitemShut {NoStop}%
\bibitem [{\citenamefont {McCumber}(1968)}]{mccumber}%
  \BibitemOpen
  \bibfield  {author} {\bibinfo {author} {\bibfnamefont {D.~E.}\ \bibnamefont
  {McCumber}},\ }\href@noop {} {\bibfield  {journal} {\bibinfo  {journal} {J.
  Appl. Phys.}\ }\textbf {\bibinfo {volume} {39}},\ \bibinfo {pages} {3113}
  (\bibinfo {year} {1968})}\BibitemShut {NoStop}%
\bibitem [{\citenamefont {Marino}\ \emph {et~al.}(1999)\citenamefont {Marino},
  \citenamefont {Raghavan}, \citenamefont {Fantoni}, \citenamefont {Shenoy},\
  and\ \citenamefont {Smerzi}}]{marino}%
  \BibitemOpen
  \bibfield  {author} {\bibinfo {author} {\bibfnamefont {I.}~\bibnamefont
  {Marino}}, \bibinfo {author} {\bibfnamefont {S.}~\bibnamefont {Raghavan}},
  \bibinfo {author} {\bibfnamefont {S.}~\bibnamefont {Fantoni}}, \bibinfo
  {author} {\bibfnamefont {S.~R.}\ \bibnamefont {Shenoy}}, \ and\ \bibinfo
  {author} {\bibfnamefont {A.}~\bibnamefont {Smerzi}},\ }\href@noop {}
  {\bibfield  {journal} {\bibinfo  {journal} {Phys. Rev. A}\ }\textbf {\bibinfo
  {volume} {60}},\ \bibinfo {pages} {487} (\bibinfo {year} {1999})}\BibitemShut
  {NoStop}%
\bibitem [{\citenamefont {Wright}\ \emph {et~al.}(2000)\citenamefont {Wright},
  \citenamefont {Arlt},\ and\ \citenamefont {Dholakia}}]{lag}%
  \BibitemOpen
  \bibfield  {author} {\bibinfo {author} {\bibfnamefont {E.~M.}\ \bibnamefont
  {Wright}}, \bibinfo {author} {\bibfnamefont {J.}~\bibnamefont {Arlt}}, \ and\
  \bibinfo {author} {\bibfnamefont {K.}~\bibnamefont {Dholakia}},\ }\href@noop
  {} {\bibfield  {journal} {\bibinfo  {journal} {Phys. Rev. A}\ }\textbf
  {\bibinfo {volume} {63}},\ \bibinfo {pages} {013608} (\bibinfo {year}
  {2000})}\BibitemShut {NoStop}%
\bibitem [{\citenamefont {Castin}\ and\ \citenamefont {Dum}(1999)}]{castin}%
  \BibitemOpen
  \bibfield  {author} {\bibinfo {author} {\bibfnamefont {Y.}~\bibnamefont
  {Castin}}\ and\ \bibinfo {author} {\bibfnamefont {R.}~\bibnamefont {Dum}},\
  }\href@noop {} {\bibfield  {journal} {\bibinfo  {journal} {Eur. Phys. J. D}\
  }\textbf {\bibinfo {volume} {7}},\ \bibinfo {pages} {399} (\bibinfo {year}
  {1999})}\BibitemShut {NoStop}%
\bibitem [{\citenamefont {Muruganandam}\ and\ \citenamefont
  {Adhikari}(2009)}]{adhikari}%
  \BibitemOpen
  \bibfield  {author} {\bibinfo {author} {\bibfnamefont {P.}~\bibnamefont
  {Muruganandam}}\ and\ \bibinfo {author} {\bibfnamefont {S.~K.}\ \bibnamefont
  {Adhikari}},\ }\href@noop {} {\bibfield  {journal} {\bibinfo  {journal}
  {Comput. Phys. Commun.}\ }\textbf {\bibinfo {volume} {180}},\ \bibinfo
  {pages} {1888} (\bibinfo {year} {2009})}\BibitemShut {NoStop}%
\bibitem [{\citenamefont {Cataldo}\ and\ \citenamefont {Jezek}(2011)}]{cat11}%
  \BibitemOpen
  \bibfield  {author} {\bibinfo {author} {\bibfnamefont {H.~M.}\ \bibnamefont
  {Cataldo}}\ and\ \bibinfo {author} {\bibfnamefont {D.~M.}\ \bibnamefont
  {Jezek}},\ }\href@noop {} {\bibfield  {journal} {\bibinfo  {journal} {Phys.
  Rev. A}\ }\textbf {\bibinfo {volume} {84}},\ \bibinfo {pages} {013602}
  (\bibinfo {year} {2011})}\BibitemShut {NoStop}%
\bibitem [{\citenamefont {Dutta}\ \emph {et~al.}(2015)\citenamefont {Dutta},
  \citenamefont {Gajda}, \citenamefont {Hauke}, \citenamefont {Lewenstein},
  \citenamefont {{\relax D.-S. L\"uhmann}}, \citenamefont {Malomed},
  \citenamefont {{\relax T. Sowi\'nski}},\ and\ \citenamefont
  {Zakrzewski}}]{duttar}%
  \BibitemOpen
  \bibfield  {author} {\bibinfo {author} {\bibfnamefont {O.}~\bibnamefont
  {Dutta}}, \bibinfo {author} {\bibfnamefont {M.}~\bibnamefont {Gajda}},
  \bibinfo {author} {\bibfnamefont {P.}~\bibnamefont {Hauke}}, \bibinfo
  {author} {\bibfnamefont {M.}~\bibnamefont {Lewenstein}}, \bibinfo {author}
  {\bibnamefont {{\relax D.-S. L\"uhmann}}}, \bibinfo {author} {\bibfnamefont
  {B.~A.}\ \bibnamefont {Malomed}}, \bibinfo {author} {\bibnamefont {{\relax T.
  Sowi\'nski}}}, \ and\ \bibinfo {author} {\bibfnamefont {J.}~\bibnamefont
  {Zakrzewski}},\ }\href@noop {} {\bibfield  {journal} {\bibinfo  {journal}
  {Rep. Prog. Phys.}\ }\textbf {\bibinfo {volume} {78}},\ \bibinfo {pages}
  {066001} (\bibinfo {year} {2015})}\BibitemShut {NoStop}%
\bibitem [{\citenamefont {Clarke}\ \emph {et~al.}(1988)\citenamefont {Clarke},
  \citenamefont {Cleland}, \citenamefont {Devoret}, \citenamefont {Esteve},\
  and\ \citenamefont {Martinis}}]{clarke}%
  \BibitemOpen
  \bibfield  {author} {\bibinfo {author} {\bibfnamefont {J.}~\bibnamefont
  {Clarke}}, \bibinfo {author} {\bibfnamefont {A.}~\bibnamefont {Cleland}},
  \bibinfo {author} {\bibfnamefont {M.}~\bibnamefont {Devoret}}, \bibinfo
  {author} {\bibfnamefont {D.}~\bibnamefont {Esteve}}, \ and\ \bibinfo {author}
  {\bibfnamefont {J.}~\bibnamefont {Martinis}},\ }\href@noop {} {\bibfield
  {journal} {\bibinfo  {journal} {Science}\ }\textbf {\bibinfo {volume}
  {239}},\ \bibinfo {pages} {992} (\bibinfo {year} {1988})}\BibitemShut
  {NoStop}%
\bibitem [{\citenamefont {Blackburn}\ \emph {et~al.}(2016)\citenamefont
  {Blackburn}, \citenamefont {Cirillo},\ and\ \citenamefont {{\relax N. Gr\o
  nbech-Jensen}}}]{blackburn}%
  \BibitemOpen
  \bibfield  {author} {\bibinfo {author} {\bibfnamefont {J.~A.}\ \bibnamefont
  {Blackburn}}, \bibinfo {author} {\bibfnamefont {M.}~\bibnamefont {Cirillo}},
  \ and\ \bibinfo {author} {\bibnamefont {{\relax N. Gr\o nbech-Jensen}}},\
  }\href@noop {} {\bibfield  {journal} {\bibinfo  {journal} {Phys. Rep.}\
  }\textbf {\bibinfo {volume} {611}},\ \bibinfo {pages} {1} (\bibinfo {year}
  {2016})}\BibitemShut {NoStop}%
\bibitem [{\citenamefont {Kwon}\ \emph {et~al.}(2019)\citenamefont {Kwon},
  \citenamefont {{\relax G. Del Pace}}, \citenamefont {Panza}, \citenamefont
  {Inguscio}, \citenamefont {Zwerger}, \citenamefont {Zaccanti}, \citenamefont
  {Scazza},\ and\ \citenamefont {Roati}}]{scazza}%
  \BibitemOpen
  \bibfield  {author} {\bibinfo {author} {\bibfnamefont {W.~J.}\ \bibnamefont
  {Kwon}}, \bibinfo {author} {\bibnamefont {{\relax G. Del Pace}}}, \bibinfo
  {author} {\bibfnamefont {R.}~\bibnamefont {Panza}}, \bibinfo {author}
  {\bibfnamefont {M.}~\bibnamefont {Inguscio}}, \bibinfo {author}
  {\bibfnamefont {W.}~\bibnamefont {Zwerger}}, \bibinfo {author} {\bibfnamefont
  {M.}~\bibnamefont {Zaccanti}}, \bibinfo {author} {\bibfnamefont
  {F.}~\bibnamefont {Scazza}}, \ and\ \bibinfo {author} {\bibfnamefont
  {G.}~\bibnamefont {Roati}},\ }\href@noop {} {} (\bibinfo {year} {2019}),\
  \Eprint {http://arxiv.org/abs/1908.09696 [cond-mat.quant-gas]}
  {arXiv:1908.09696 [cond-mat.quant-gas]} \BibitemShut {NoStop}%
\bibitem [{\citenamefont {Barnett}\ and\ \citenamefont
  {Vaccaro}(2007)}]{barnett}%
  \BibitemOpen
  \bibinfo {editor} {\bibfnamefont {S.~M.}\ \bibnamefont {Barnett}}\ and\
  \bibinfo {editor} {\bibfnamefont {J.~A.}\ \bibnamefont {Vaccaro}},\ eds.,\
  \href@noop {} {\emph {\bibinfo {title} {The Quantum Phase Operator: A
  Review}}}\ (\bibinfo  {publisher} {Taylor \& Francis},\ \bibinfo {address}
  {New York},\ \bibinfo {year} {2007})\BibitemShut {NoStop}%
\bibitem [{\citenamefont {Higham}(2008)}]{higham}%
  \BibitemOpen
  \bibfield  {author} {\bibinfo {author} {\bibfnamefont {N.~J.}\ \bibnamefont
  {Higham}},\ }\href@noop {} {\emph {\bibinfo {title} {Functions of Matrices:
  Theory and Computation}}}\ (\bibinfo  {publisher} {SIAM},\ \bibinfo {address}
  {Philadelphia},\ \bibinfo {year} {2008})\BibitemShut {NoStop}%
\bibitem [{\citenamefont {Susskind}\ and\ \citenamefont
  {Glogower}(1964)}]{suss}%
  \BibitemOpen
  \bibfield  {author} {\bibinfo {author} {\bibfnamefont {L.}~\bibnamefont
  {Susskind}}\ and\ \bibinfo {author} {\bibfnamefont {J.}~\bibnamefont
  {Glogower}},\ }\href@noop {} {\bibfield  {journal} {\bibinfo  {journal}
  {Physics}\ }\textbf {\bibinfo {volume} {1}},\ \bibinfo {pages} {49} (\bibinfo
  {year} {1964})}\BibitemShut {NoStop}%
\bibitem [{\citenamefont {Mehta}\ \emph {et~al.}(1992)\citenamefont {Mehta},
  \citenamefont {Roy},\ and\ \citenamefont {Saxena}}]{saxena}%
  \BibitemOpen
  \bibfield  {author} {\bibinfo {author} {\bibfnamefont {C.~L.}\ \bibnamefont
  {Mehta}}, \bibinfo {author} {\bibfnamefont {A.~K.}\ \bibnamefont {Roy}}, \
  and\ \bibinfo {author} {\bibfnamefont {G.~M.}\ \bibnamefont {Saxena}},\
  }\href@noop {} {\bibfield  {journal} {\bibinfo  {journal} {Phys. Rev. A}\
  }\textbf {\bibinfo {volume} {46}},\ \bibinfo {pages} {1565} (\bibinfo {year}
  {1992})}\BibitemShut {NoStop}%
\bibitem [{\citenamefont {Transtrum}\ and\ \citenamefont {{\relax J.-F. S. Van
  Huele}}(2005)}]{huele}%
  \BibitemOpen
  \bibfield  {author} {\bibinfo {author} {\bibfnamefont {M.~K.}\ \bibnamefont
  {Transtrum}}\ and\ \bibinfo {author} {\bibnamefont {{\relax J.-F. S. Van
  Huele}}},\ }\href@noop {} {\bibfield  {journal} {\bibinfo  {journal} {J.
  Math. Phys.}\ }\textbf {\bibinfo {volume} {46}},\ \bibinfo {pages} {063510}
  (\bibinfo {year} {2005})}\BibitemShut {NoStop}%
\bibitem [{\citenamefont {Golubov}\ \emph {et~al.}(2004)\citenamefont
  {Golubov}, \citenamefont {Kupriyanov},\ and\ \citenamefont {{\relax E.
  Il'ichev}}}]{golubov}%
  \BibitemOpen
  \bibfield  {author} {\bibinfo {author} {\bibfnamefont {A.~A.}\ \bibnamefont
  {Golubov}}, \bibinfo {author} {\bibfnamefont {M.~Y.}\ \bibnamefont
  {Kupriyanov}}, \ and\ \bibinfo {author} {\bibnamefont {{\relax E.
  Il'ichev}}},\ }\href@noop {} {\bibfield  {journal} {\bibinfo  {journal} {Rev.
  Mod. Phys.}\ }\textbf {\bibinfo {volume} {76}},\ \bibinfo {pages} {411}
  (\bibinfo {year} {2004})}\BibitemShut {NoStop}%
\bibitem [{\citenamefont {Mueller}(2002)}]{mueller}%
  \BibitemOpen
  \bibfield  {author} {\bibinfo {author} {\bibfnamefont {E.~J.}\ \bibnamefont
  {Mueller}},\ }\href@noop {} {\bibfield  {journal} {\bibinfo  {journal} {Phys.
  Rev. A}\ }\textbf {\bibinfo {volume} {66}},\ \bibinfo {pages} {063603}
  (\bibinfo {year} {2002})}\BibitemShut {NoStop}%
\bibitem [{\citenamefont {Xhani}\ \emph {et~al.}(2020)\citenamefont {Xhani},
  \citenamefont {Neri}, \citenamefont {Galantucci}, \citenamefont {Scazza},
  \citenamefont {Burchianti}, \citenamefont {{\relax K.-L. Lee}}, \citenamefont
  {Barenghi}, \citenamefont {Trombettoni}, \citenamefont {Inguscio},
  \citenamefont {Zaccanti}, \citenamefont {Roati},\ and\ \citenamefont
  {Proukakis}}]{xhani}%
  \BibitemOpen
  \bibfield  {author} {\bibinfo {author} {\bibfnamefont {K.}~\bibnamefont
  {Xhani}}, \bibinfo {author} {\bibfnamefont {E.}~\bibnamefont {Neri}},
  \bibinfo {author} {\bibfnamefont {L.}~\bibnamefont {Galantucci}}, \bibinfo
  {author} {\bibfnamefont {F.}~\bibnamefont {Scazza}}, \bibinfo {author}
  {\bibfnamefont {A.}~\bibnamefont {Burchianti}}, \bibinfo {author}
  {\bibnamefont {{\relax K.-L. Lee}}}, \bibinfo {author} {\bibfnamefont
  {C.~F.}\ \bibnamefont {Barenghi}}, \bibinfo {author} {\bibfnamefont
  {A.}~\bibnamefont {Trombettoni}}, \bibinfo {author} {\bibfnamefont
  {M.}~\bibnamefont {Inguscio}}, \bibinfo {author} {\bibfnamefont
  {M.}~\bibnamefont {Zaccanti}}, \bibinfo {author} {\bibfnamefont
  {G.}~\bibnamefont {Roati}}, \ and\ \bibinfo {author} {\bibfnamefont {N.~P.}\
  \bibnamefont {Proukakis}},\ }\href@noop {} {\bibfield  {journal} {\bibinfo
  {journal} {Phys. Rev. Lett.}\ }\textbf {\bibinfo {volume} {124}},\ \bibinfo
  {pages} {045301} (\bibinfo {year} {2020})}\BibitemShut {NoStop}%
\bibitem [{\citenamefont {Amico}\ \emph {et~al.}(2017)\citenamefont {Amico},
  \citenamefont {Birkl}, \citenamefont {Boshier},\ and\ \citenamefont {{\relax
  L.-C. Kwek}}}]{amico17}%
  \BibitemOpen
  \bibfield  {author} {\bibinfo {author} {\bibfnamefont {L.}~\bibnamefont
  {Amico}}, \bibinfo {author} {\bibfnamefont {G.}~\bibnamefont {Birkl}},
  \bibinfo {author} {\bibfnamefont {M.}~\bibnamefont {Boshier}}, \ and\
  \bibinfo {author} {\bibnamefont {{\relax L.-C. Kwek}}},\ }\href@noop {}
  {\bibfield  {journal} {\bibinfo  {journal} {New. J. Phys.}\ }\textbf
  {\bibinfo {volume} {19}},\ \bibinfo {pages} {020201} (\bibinfo {year}
  {2017})}\BibitemShut {NoStop}%
\bibitem [{\citenamefont {Lee}\ \emph {et~al.}(2013)\citenamefont {Lee},
  \citenamefont {McIlvain}, \citenamefont {Lobb},\ and\ \citenamefont {{\relax
  W. T. Hill, III}}}]{lee}%
  \BibitemOpen
  \bibfield  {author} {\bibinfo {author} {\bibfnamefont {J.~G.}\ \bibnamefont
  {Lee}}, \bibinfo {author} {\bibfnamefont {B.~J.}\ \bibnamefont {McIlvain}},
  \bibinfo {author} {\bibfnamefont {C.~J.}\ \bibnamefont {Lobb}}, \ and\
  \bibinfo {author} {\bibnamefont {{\relax W. T. Hill, III}}},\ }\href@noop {}
  {\bibfield  {journal} {\bibinfo  {journal} {Sci. Rep.}\ }\textbf {\bibinfo
  {volume} {3}},\ \bibinfo {pages} {1034} (\bibinfo {year} {2013})}\BibitemShut
  {NoStop}%
\bibitem [{\citenamefont {Eckel}\ \emph {et~al.}(2016)\citenamefont {Eckel},
  \citenamefont {Lee}, \citenamefont {Jendrzejewski}, \citenamefont {Lobb},
  \citenamefont {Campbell},\ and\ \citenamefont {{\relax W. T. Hill,
  III}}}]{eckel16}%
  \BibitemOpen
  \bibfield  {author} {\bibinfo {author} {\bibfnamefont {S.}~\bibnamefont
  {Eckel}}, \bibinfo {author} {\bibfnamefont {J.~G.}\ \bibnamefont {Lee}},
  \bibinfo {author} {\bibfnamefont {F.}~\bibnamefont {Jendrzejewski}}, \bibinfo
  {author} {\bibfnamefont {C.~J.}\ \bibnamefont {Lobb}}, \bibinfo {author}
  {\bibfnamefont {G.~K.}\ \bibnamefont {Campbell}}, \ and\ \bibinfo {author}
  {\bibnamefont {{\relax W. T. Hill, III}}},\ }\href@noop {} {\bibfield
  {journal} {\bibinfo  {journal} {Phys. Rev. A}\ }\textbf {\bibinfo {volume}
  {93}},\ \bibinfo {pages} {063619} (\bibinfo {year} {2016})}\BibitemShut
  {NoStop}%
\bibitem [{\citenamefont {Burchianti}\ \emph {et~al.}(2018)\citenamefont
  {Burchianti}, \citenamefont {Scazza}, \citenamefont {Amico}, \citenamefont
  {Valtolina}, \citenamefont {Seman}, \citenamefont {Fort}, \citenamefont
  {Zaccanti}, \citenamefont {Inguscio},\ and\ \citenamefont
  {Roati}}]{burchianti}%
  \BibitemOpen
  \bibfield  {author} {\bibinfo {author} {\bibfnamefont {A.}~\bibnamefont
  {Burchianti}}, \bibinfo {author} {\bibfnamefont {F.}~\bibnamefont {Scazza}},
  \bibinfo {author} {\bibfnamefont {A.}~\bibnamefont {Amico}}, \bibinfo
  {author} {\bibfnamefont {G.}~\bibnamefont {Valtolina}}, \bibinfo {author}
  {\bibfnamefont {J.~A.}\ \bibnamefont {Seman}}, \bibinfo {author}
  {\bibfnamefont {C.}~\bibnamefont {Fort}}, \bibinfo {author} {\bibfnamefont
  {M.}~\bibnamefont {Zaccanti}}, \bibinfo {author} {\bibfnamefont
  {M.}~\bibnamefont {Inguscio}}, \ and\ \bibinfo {author} {\bibfnamefont
  {G.}~\bibnamefont {Roati}},\ }\href@noop {} {\bibfield  {journal} {\bibinfo
  {journal} {Phys. Rev. Lett.}\ }\textbf {\bibinfo {volume} {120}},\ \bibinfo
  {pages} {025302} (\bibinfo {year} {2018})}\BibitemShut {NoStop}%
\bibitem [{\citenamefont {Martinis}(2004)}]{martinis}%
  \BibitemOpen
  \bibfield  {author} {\bibinfo {author} {\bibfnamefont {J.~M.}\ \bibnamefont
  {Martinis}},\ }\href@noop {} {\bibfield  {journal} {\bibinfo  {journal} {Les
  Houches}\ }\textbf {\bibinfo {volume} {79}},\ \bibinfo {pages} {487}
  (\bibinfo {year} {2004})}\BibitemShut {NoStop}%
\bibitem [{\citenamefont {Bidasyuk}\ \emph {et~al.}(2018)\citenamefont
  {Bidasyuk}, \citenamefont {Weyrauch}, \citenamefont {Momme},\ and\
  \citenamefont {Prikhodko}}]{bidasyuk}%
  \BibitemOpen
  \bibfield  {author} {\bibinfo {author} {\bibfnamefont {Y.~M.}\ \bibnamefont
  {Bidasyuk}}, \bibinfo {author} {\bibfnamefont {M.}~\bibnamefont {Weyrauch}},
  \bibinfo {author} {\bibfnamefont {M.}~\bibnamefont {Momme}}, \ and\ \bibinfo
  {author} {\bibfnamefont {O.~O.}\ \bibnamefont {Prikhodko}},\ }\href@noop {}
  {\bibfield  {journal} {\bibinfo  {journal} {J. Phys. B: At. Mol. Opt. Phys.}\
  }\textbf {\bibinfo {volume} {51}},\ \bibinfo {pages} {205301} (\bibinfo
  {year} {2018})}\BibitemShut {NoStop}%
\end{thebibliography}

\providecommand{\noopsort}[1]{}\providecommand{\singleletter}[1]{#1}%

\end{document}